\documentclass[journal,onecolumn,draftclsnofoot,fleqn]{IEEEtran1}
\usepackage{multicol}
\usepackage{amsmath}
\usepackage{graphicx}
\usepackage{epsfig}
\usepackage{psfrag}
\usepackage[nomarkers]{endfloat}
\usepackage{amsfonts}
\usepackage{epstopdf}
\usepackage{amssymb}
\usepackage{cite}
\usepackage{amsthm,amssymb}

\begin{document}

\title{Stable Throughput in a Cognitive Wireless Network}

\author{Anthony Fanous,~\IEEEmembership{Student Member,~IEEE}
        and Anthony Ephremides,~\IEEEmembership{Fellow,~IEEE}
\thanks{The authors are with the Department of Electrical and Computer Engineering, University of Maryland, College Park, MD, 20742, USA (e-mail: afanous@umd.edu;
etony@umd.edu).}}

\maketitle
\begin{abstract}
\textbf{We study, from a network layer perspective, the effect of an Ad-Hoc secondary network with $N$ nodes accessing the spectrum licensed to a primary node. Specifically, a network consisting of one primary source-destination pair and $N$ secondary cognitive source-destination pairs that randomly access the channel during the idle slots of the primary user is considered. Both cases of perfect and imperfect sensing are considered and we adopt the SINR threshold model for detection to properly model the interference throughout the network. We first study the effect of the number of secondary nodes as well as secondary nodes' transmission parameters such as the power and the channel access probabilities on the stable throughput of the primary node. If the sensing is perfect, then the secondary nodes do not interfere with the primary node and thus do not affect its stable throughput. However, if the sensing is imperfect, then the secondary nodes should control their interference on the primary node in order to keep its queue stable. It is shown that if the primary user's arrival rate is less than some calculated finite value, cognitive nodes can employ any transmission power or probabilities without affecting the primary user's stability; otherwise, the secondary nodes should control their transmission parameters to reduce the interference on the primary. It is also shown that in contrast with the primary maximum stable throughput which strictly decreases with increased sensing errors, the throughput of the secondary nodes might increase with sensing errors as more transmission opportunities become available to them. Finally, we explore the use of the secondary nodes as relays of the primary node's traffic to compensate for the interference they might cause. Specifically, we introduce a relaying protocol based on distributed space-time coding that forces all the secondary nodes that are able to decode a primary's unsuccessful packet to relay that packet whenever the primary is idle. In this case, for appropriate modulation scheme and under perfect sensing, it is shown that the more secondary nodes in the system, the better for the primary user in terms of his stable throughput. Meanwhile, the secondary nodes might benefit from relaying by having access to a larger number of idle slots becoming available to them due to the increase of the service rate of the primary. For the case of a single secondary node, the proposed relaying protocol guarantees that either both the primary and the secondary benefit from relaying or none of them does.}
\end{abstract}
\IEEEpeerreviewmaketitle
\section{Introduction}
Restricting the spectrum access only to licensed users represents a highly inefficient resource utilization as actual measurements indicated that such resources remain idle for long proportions of time \cite{FCC_1, Measurement_1, Measurement_2}. This observation as well as the development of sophisticated nodes capable of exploring licensed spectrum and adjusting their transmission parameters accordingly, motivated the idea of cognitive radio \cite{Mitola_PhD,Haykin_1,Next_Survey,liang_survey} where the spectrum is made available to both licensed (also called primary) users as well as unlicensed (secondary/cognitive) users who opportunistically access the spectrum in such a way that the interference on the primary users is limited or even completely avoided.

Several approaches to cognitive radio operations have been suggested in the literature \cite{Natasha_Magazine,Jafar_Mag,Next_Survey}. Two main paradigms exist for cognitive access, namely, spectrum sharing (SS) and opportunistic spectrum access (OSA). In spectrum sharing systems, secondary users are allowed to transmit concurrently with the primary users given some measures to keep the interference caused on primary users within an allowable range (usually within the primary node's noise floor). Opportunistic spectrum access systems aim at avoiding concurrent transmissions between primary and secondary users by restricting the secondary users to access the channel only at unoccupied temporal, spectral or spatial holes. In order to achieve such goal, secondary users have to sense the channel at every slot to identify whether a primary transmission is ongoing or not\cite{beacon,Sensing_Survey,Sensing_2}. Many cooperative sensing techniques have also been suggested \cite{Coop_Sensing} which efficiently mitigate the hidden terminal problem.

Cooperative communications was motivated by the effectiveness of space diversity in combatting fading, and hence single antenna users can benefit by the virtual MIMO effect induced by other nodes relaying their transmissions. Single relay channel has been studied from an information theoretic point of view in \cite{Cover_ElGamal}, but Shannon capacity was exactly characterized only for the case of physically degraded channels. The relay node in these works was dedicated to forward the source message, i.e., it does not have its own traffic. Later, cooperative protocols for two sources- two destinations case have been proposed and analyzed in \cite{Laneman_1} and distributed space-time codes for multiple relay scenarios have been developed in \cite{Distributed_ST_1,Distributed_ST_2}. However, the performance was based on information theoretic metrics such as capacity regions, achievable rates and outage probabilities. A network-level cooperative protocol for an uplink where a single pure cognitive relay is introduced to forward unsuccessful packets from source nodes during their idle slots has been proposed and analyzed in \cite{Sadek_1} with stable throughput and average delay as performance metrics under the assumption of perfect sensing. The assumption of pure relay has been relaxed in \cite{Beiyu_1,Beiyu_2} where the relay node is a source node having its own traffic but multi-relay case was not considered.

Cognitive radio has been studied from an information theoretic point of view in \cite{Natasha_IT,Vishwanath_IT}. However, such formulation does not take into account the bursty nature of the traffic and mainly focuses on sophisticated coding techniques at the physical layer for relaying and interference nulling rather than network layer aspects such as stable throughput and delay analysis. The stable throughput of simple cognitive networks has been studied in In \cite{Simeone_NJIT,Sastry_1,krikidis_2}. In \cite{Simeone_NJIT}, authors studied the stable throughput of a simple cognitive network consisting of one primary and one secondary source-destination pairs under the SINR threshold model for reception with and without relaying for perfect and erroneous sensing. However, such formulation does not capture the effect of the number of secondary nodes on the performance. Moreover, many simplifying assumptions were made such as neglecting the false alarm probability (which has huge impact for more than one secondary node as we will see later) and ignoring the case where secondary node can be successful when the primary transmits simultaneously. By having one secondary node, cooperation is restricted to single node relaying which is beneficial only if the average received SNR on the secondary source-primary destination link is larger than the average SNR on the primary source-primary destination link which is not guaranteed especially if the secondary source has to reduce its power to protect the primary from the interference due to sensing errors. This weakness is overcome in our work by using multinode relaying. In \cite{Sastry_1}, a similar model was considered under general probabilistic reception model where the secondary node accesses the channel with probability one if sensed to be idle and with some optimized probability $p$ if sensed to be busy. In \cite{krikidis_2}, a multiaccess (MAC) primary network with two primary transmitters with a single secondary node is considered. In \cite{ElSherif1}, authors considered a more realistic model of a primary network consisting of a TDMA uplink with some dedicated cognitive relays deployed to help the primary, and secondary network consisting of an Ad-Hoc network with stable throughput as performance criterion. However, only the case of perfect sensing of the primary nodes was considered which is an idealistic assumption far from reality. The model for reception used is the collision channel model which is not a realistic assumption for currently used sophisticated receivers with multipacket reception (MPR) capability \cite{Tong_MPR}; and in specific, it is not realistic for Ad-Hoc networks where different source-destination pairs can be largely separated in space and thus the interference of one on the other when transmitting simultaneously is likely to be small. Furthermore, only one relay helps the primary source node at a time, i.e., no multinode relaying was considered despite the presence of several dedicated relays in the system which largely limits the potential gain of relaying. In \cite{Krikidis_1}, the stable throughput of a network consisting of one primary link and a symmetric secondary cluster with physical layer enhancements and perfect sensing is considered. The secondary cluster is controlled via a central controller and communication within the cluster is assumed to be perfect with no overhead and only one secondary node is scheduled at each slot. Single node relaying has also been studied. However, the assumption of having a secondary cluster is not appropriate for Ad-Hoc networks where the presence of a central controller is not generally feasible and the secondary transmissions interfere.

This work is motivated by the fact that in a typical cognitive scenario, secondary users sense different licensed frequency bands to capture possible opportunities. We study from a network layer perspective via a cross-layer analysis the effect of a secondary Ad-Hoc network with $N$ transmitter-receiver pairs sharing the licensed band of a primary node on the stable throughput of the primary node. The throughput of the secondary network and the possibility of multinode relaying of the primary's unsuccessful packets are also considered. We consider a primary network consisting of a single point to point link and a secondary network consisting of $N$ source-destination pairs that randomly access the channel. We adopt the physical interference model for reception (SINR threshold model) which captures the possibility of MPR capability in contrast with the oversimplified collision model. We study the effect of the interference of the secondary nodes on the primary node's stable throughput rate, which may occur due to incorrect sensing or even with perfect sensing in the presence of malicious attacks\cite{security}. Secondary queues are assumed to be saturated to avoid queueing interaction which is known to be a thorny problem and stability region is exactly known only for two and three nodes \cite{Rao,Luo,Szpankowski}. We then study the maximization of the secondary nodes' sum throughput over the feasible set of transmission power and channel access probabilities that guarantees the primary protection. Having shown the detrimental effect of a large number of secondary nodes interfering with the primary because of imperfect sensing, we propose a relaying protocol that utilizes all the secondary nodes that can decode a primary unsuccessful packet to relay that packet by using orthogonal space-time block codes \cite{Tarokh_1,Tarokh_2,ST_3}. It is shown that under this protocol, the more secondary nodes present in the network, the more the primary node benefits in terms of its maximum stable throughput. Meanwhile, the secondary nodes might benefit from such relaying. The primary node benefits by having more nodes relaying its packets and the secondary might benefit by helping the primary emptying his queue and hence having access to a larger number of idle slots. It is also shown that for a network with a single secondary node, the proposed protocol guarantees that either both the primary and the secondary nodes benefit from relaying or none of them benefits.

The paper is organized as follows: In section II, we describe the network and channel models. In section III, we introduce basic definitions and theorems concerning the queue stability that will be used throughout the paper. Section IV studies the throughputs of the primary and secondary nodes in the perfect sensing case which serves as an upper bound on the performance, while, in section V, we analyze the effect of erroneous sensing on the primary's and secondary's throughputs. In section VI, we propose and analyze the relaying protocol to benefit of the large population of secondary nodes. Section VII presents the numerical results and in section VIII, we conclude the paper.

\section{System Model}
\begin{figure}
\centering
\includegraphics[width=1\textwidth]{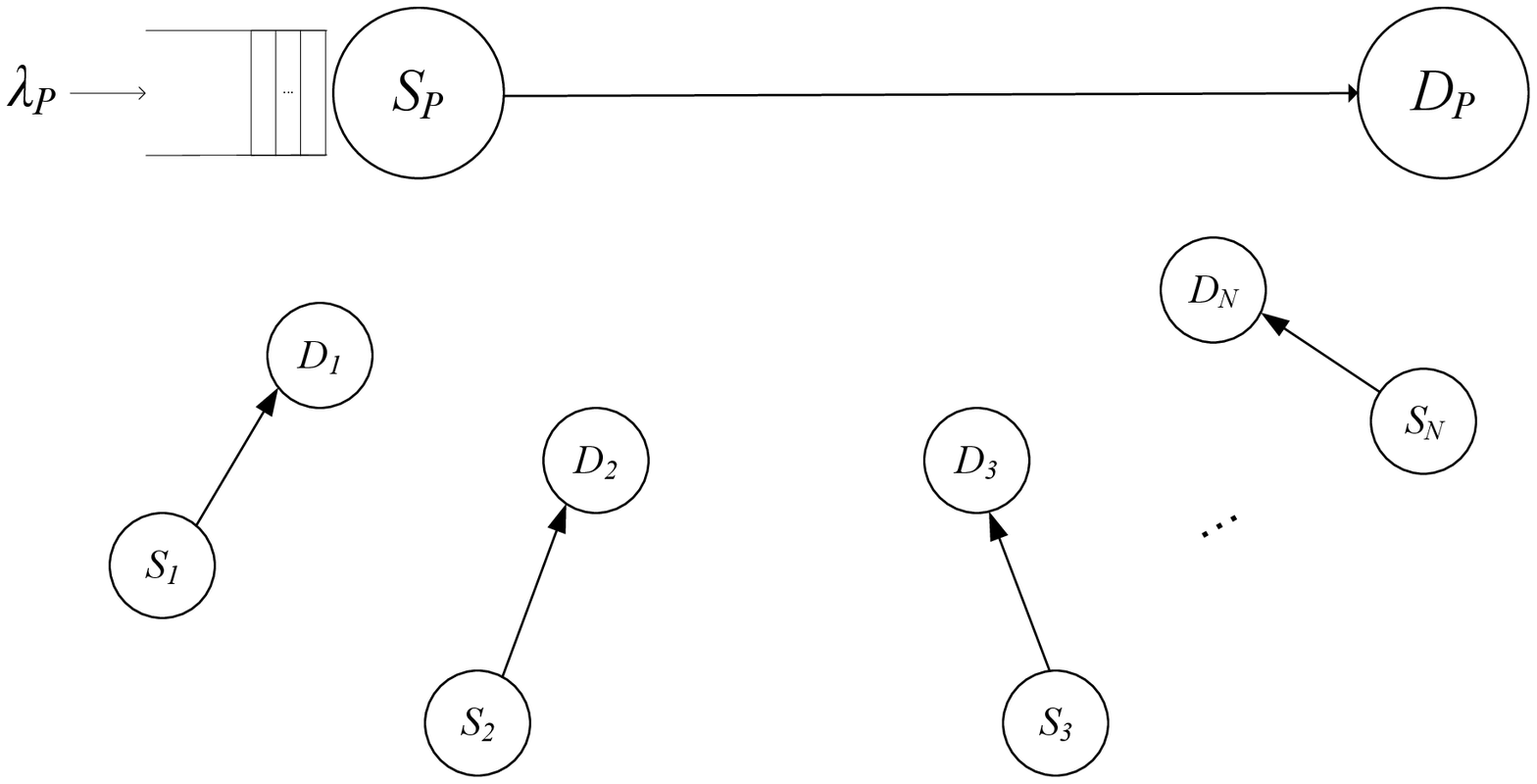}
\caption{System Model}\label{Fig0}
\end{figure}
The system consists of one primary link and a secondary network consisting of \textit{N} secondary cognitive source-destination pairs forming an interference network as shown in Figure \ref{Fig0}. All the nodes use the same frequency band for transmission. This situation arises when that band is licensed to the primary node, while to improve the spectral efficiency, some secondary nodes are allowed to access that band in an opportunistic way. All nodes have buffers of infinite capacity to store their packets to be transmitted. Time is slotted with one packet duration equal to the slot duration. Arrival process to the primary source node is assumed to be stationary with an average rate $\lambda_P$ packets/slot, while secondary source nodes are assumed to be saturated. Throughout the paper, we designate the primary node by the subscript \textit{P} and the \textit{i}th secondary node by the subscript \textit{i} with $i\in\{1,2,...,N\}$. The $i$th source node is denoted by $S_i$ and the $i$th destination node is denoted by $D_i$, $i\in\{P,1,2,...,N\}$. The \textit{i}th source node transmits at power $P_i$, $i\in\{P,1,2,...,N\}$.
\subsection{Channel Model:}
 The distance between node $i$ and node $j$ is denoted by $r_{ij}$, where $i,j\in\{S_k,D_k|k=P,1,2,...,N\}$. For instance, $r_{S_PD_j}$ denotes the distance between the primary source and the $j$th secondary destination node. Path loss exponent is assumed to be equal to ($\alpha$) throughout the network. The link between the $(i,j)$ pair of nodes is subject to stationary block fading with fading coefficients $h_{ij}$ which are independent over slots and mutually independent among links. All nodes are subject to independent additive white complex Gaussian noise with zero mean and variance $N_0$. Under the adopted $SINR$ threshold model for reception, node $j$ is able to successfully decode a packet if the received signal-to-interference plus noise ratio (SINR) remains larger than some threshold $\beta_j$ throughout the packet duration \cite{Int_Modeling}. The threshold $\beta_j$ depends on the modulation scheme, the coding and the target BER set by the receiving node as well as other features of the detector structure. Upon the success or failure of a packet reception at a node, an Acknowledgment/Non-Acknowledgment (ACK/NACK) packet is fedback to the corresponding transmitter. The ACK/NACK packets are assumed to be instantaneous and error free which is a reasonable assumption for short length ACK/NACK packets that have negligible delay, and small error rate achieved by using low rate codes on the feedback channel.\\
Under this model, the received signal at the $j$th node of the transmitted signal by the $i$th node in the presence of an interfering set of nodes $\mathcal{I}$ at time slot $(t)$ is given by:
\begin{equation}
y_j^t=\sqrt{P_ir_{ij}^{-\alpha}}h_{ij}^tx_i^t+\sum_{k\in\mathcal{I}}\sqrt{P_kr_{kj}^{-\alpha}}h_{kj}^tx_k^t+n_j^t\label{1}
\end{equation}
where $x_k^t$ is the transmitted packet by the $k$th node at time slot $(t)$ and is of unit power while $n_j^t \sim \mathcal{CN}(0,N_0)$ is the additive white complex Gaussian noise at node $j$.\\
In this case, the success probability of the $i$th node transmission at the receiving node $j$ is given by:
\begin{equation}
\mathrm{Pr}\left[\mathrm{SINR}_{ij}>\beta_j\right]=\mathrm{Pr}\left[\frac{P_ir_{ij}^{-\alpha}|h_{ij}^t|^2}{N_0+\sum_{k\in\mathcal{I}}P_kr_{kj}^{-\alpha}|h_{kj}^t|^2}>\beta_j\right]\label{2}
\end{equation}
\subsection{Multiple Access Protocol:}
Both the primary and the secondary users transmit over the same frequency band, and hence, secondary users are restricted to use the idle slots of the primary. The primary node has the priority for transmission. At the beginning of each slot, the secondary nodes sense the channel and only if a slot is detected to be idle, do they access the channel in a random access way. The $i$th secondary node will transmit in a slot with a probability $q_i$ whenever that slot is detected to be idle. We assume that there is sufficient guard time at the beginning of each slot to allow the secondary nodes to sense the channel.

According to the cognitive radio principle, secondary nodes should be ``transparent'' to the primary in the sense that their transmissions should not affect some performance criterion (here, the queueing stability) of the primary node. If the sensing is perfect, the secondary nodes never interfere with the primary and can employ any transmission parameters (power/channel access probability) that maximize their sum throughput without affecting the stability of primary. However, if the sensing is not perfect, the secondary nodes must limit their interference on the primary by controlling their transmission parameters to achieve that goal while maximizing their opportunistic throughput. We discuss the constraints on the secondary transmission parameters in case of imperfect sensing in section IV.

Throughout the work, we consider both perfect and imperfect sensing. Perfect sensing is an optimistic case and only serves as an upper bound on performance. We also consider both cases of asymmetric network with arbitrary fading distribution and the symmetric network with Rayleigh fading where $h_{ij}\sim\mathcal{CN}(0,\sigma_{ij}^2)$. We will find that most results apply for both cases but in the asymmetric case, some results are not in closed form while the symmetric case is easier to analyze and get intuition from.\\
The symmetric case with Rayleigh fading that we will consider is characterized as follows:
\begin{itemize}
\item $P_j=P_0$, for $j\in\{1,2,...,N\}$.
\item $q_j=q$, for $j\in\{1,2,...,N\}$.
\item $\beta_j=\beta$, for $j\in\{1,2,...,N\}$.
\item $r_{S_kD_j}=r_j$, for $j,k\in\{1,2,...,N\}$.
\item $r_{S_jD_P}=r_0$, for $j\in\{1,2,...,N\}$.
\item $r_{S_PS_k}=r$, for $k\in\{1,2,...,N\}$.
\item $h_{S_jD_k}=\tilde{h_j}\sim\mathcal{CN}(0,\tilde{\sigma}^2)$, for $j,k\in\{1,2,...,N\}$.
\item $h_{S_jD_P}=\bar{h_j}\sim\mathcal{CN}(0,\sigma_0^2)$, for $j\in\{1,2,...,N\}$.
\item $h_{S_PS_k}=h_k\sim\mathcal{CN}(0,\sigma^2)$, for $k\in\{1,2,...,N\}$.
\end{itemize}
This geometry, for instance, arises whenever the secondary sources lie on a circle and secondary destinations, along with primary source-destination pair lie on a line passing by the center of that circle and perpendicular to its plane.
\section{Queue Stability}
We adopt the definition of stability used by Szpankowski in \cite{Szpankowski}. \\
\textit{Definition 1:}\\
A multidimensional stochastic process $\mathbf{Q^t}=\mathit{(Q_1^t,...,Q_M^t)}$ is \textit{stable} if for every \textbf{x}$\in$ $\mathbb{N}_0^M$
the following holds:
\begin{equation}
\lim_{t\to\infty}\mathrm{Pr}[\mathbf{Q^t}<\mathbf{x}]=F(\mathbf{x}) \textrm{  and  } \lim_{\mathbf{x}\to\infty}F(\mathbf{x})=1\label{3}
\end{equation}
where $F(\mathbf{x})$ is the limiting distribution function and \textbf{x}$\rightarrow\infty$ is componentwise limit.\\
If a weaker condition holds, namely,
\begin{equation}
\lim_{\mathbf{x}\to\infty}\liminf_{t\to\infty}\mathrm{Pr}[\mathbf{Q^t<x}]=1\label{4}
\end{equation}
then the process is called substable.\\
Under the saturation assumption of the secondary source queues, there is no issue of stability except for the primary queue.\\
The primary source queue evolves as:
\begin{equation}
Q_P^{t+1}=\left(Q_P^t-Y_P^t\right)^++X_P^t\label{5}
\end{equation}
where:\\
$Q_P^t$ is the length of primary source queue at the beginning of time slot $t$.
$X_P^t$ and $Y_P^t$ are the arrival and the service processes at the primary source queue in time slot $t$ respectively and $(x)^+$=max$(x,0)$.\\
Throughout the paper, we use the following lemma \cite{Loynes,Szpankowski} sometimes referred to as Loynes' Theorem:\\
\textit{Lemma 1:}\\
For a queue evolving as in (\ref{5}). Let the pair $(X_P^t,Y_P^t)$ be strictly stationary process (i.e. $X_P^t$ and $Y_P^t$ are jointly stationary), then the following holds:\\
(i) If \textbf{E}$[X_P^t] <$ \textbf{E}$[Y_P^t]$, then the queue is stable in the sense of the definition in (\ref{3}).\\
(ii) If \textbf{E}$[X_P^t] >$ \textbf{E}$[Y_P^t]$, then the queue is unstable and $\lim_{t\to\infty}Q_P^t=\infty$ almost surely.\\
The arrival process at the primary source queue is stationary by assumption and is independent of its service process, hence a necessary and sufficient condition for stability of the primary source queue is that $\lambda_P <$\textbf{E}$[Y_P^t]$.
\section{Perfect Sensing Case}
In this case, the secondary nodes are able to perfectly identify the primary idle slots where they can access the channel from the busy slots where they must remain silent to avoid interfering with the primary. In this case, the primary gets its maximum possible service rate. Clearly, this is an ideal situation serving as an upper bound on the performance.
\subsection{Primary Queue}
\textit{Theorem 1:}\\
The stability condition for the primary queue in the perfect sensing case is given by:
\begin{equation}
\lambda_P<\mu_P^{\mathrm{max}}=\mathrm{Pr}\left[\frac{P_P|h_{S_PD_P}|^2r_{S_PD_P}^{-\alpha}}{N_0}>\beta_P\right]\label{6}
\end{equation}
For Rayleigh fading, we have that $\mu_P^{\max}=\exp\left(\frac{-N_0\beta_P}{P_P\sigma_{S_PD_P}^2r_{S_PD_P}^{-\alpha}}\right)$.
\begin{proof}
The service process of the primary node is given by $
Y_P^t=\textbf{1}\left\{\overline{\mathcal{O}_{S_PD_P}^t}\right\}$, where $\overline{\mathcal{O}_{S_PD_P}^t}$ denotes the event of no outage at the primary destination node in slot $t$ and \textbf{1}$\{\cdot\}$ is the indicator function which takes the value of one if its argument is true and zero otherwise. This event depends on the fading process on the $(S_P,D_P)$ link which is stationary and hence, $Y_P^t$ is stationary.\\
The average service rate of the primary queue in this case, that we denote by $\mu_P^{\mathrm{max}}$, is given by:
\begin{equation}
\mu_P^{\mathrm{max}}=E\left[Y_P^t\right]=\mathrm{Pr}\left[\overline{\mathcal{O}_{S_PD_P}^t}\right]=\mathrm{Pr}\left[\frac{P_P|h_{S_PD_P}|^2r_{S_PD_P}^{-\alpha}}{N_0}>\beta_P\right]\label{8}
\end{equation}
For the Rayleigh fading case, $|h_{S_PD_P}|^2$ is exponentially distributed with mean $\sigma_{S_PD_P}^2$, and the results follows.\\
Finally, by using Loynes' theorem (Lemma $1$), we can get the stability condition of the primary node as given in (\ref{6}).
\end{proof}
\subsection{Secondary Queues}
Source node $(i),i \in \mathcal{S}=\{1,2,...,N\}$ of the secondary cluster transmits with probability $(q_i)$- independently of the other secondary nodes- whenever a slot is detected to be idle.\\
\textit{Theorem 2:}\\
In the perfect sensing case, the throughput region of the secondary network is given by:
\begin{equation}
\mathcal{R}=\bigcup_{\text{\textbf{q}}\in[0,1]^N}\left\{\mbox{\boldmath$\lambda$}\in\mathbb{R}_+^N:\lambda_j=\left(1-\frac{\lambda_P}{\mu_P^{\mathrm{max}}}\right)\sum_{\substack{\mathcal{T}\subseteq\{1,2,...,N\}\\j\in\mathcal{T}}}\prod_{k\in\mathcal{T}}q_k\prod_{l\in\mathcal{S}\setminus\mathcal{T}}(1-q_l)P_{S_j}^{\mathcal{T}},\mbox{ } j\in\{1,2,...,N\}\right\}\label{9}
\end{equation}
where:
\begin{equation*}
P_{S_j}^{\mathcal{T}}=\mathrm{Pr}\left[\frac{P_{j}|h_{S_jD_j}|^2r_{S_jD_j}^{-\alpha}}{N_0+\sum_{k\in \mathcal{T},k\ne j}P_k|h_{S_kD_j}|^2r_{S_kD_j}^{-\alpha}}>\beta_{j}\right]
\end{equation*}
which is -for the Rayleigh fading case- equal to:
\begin{equation*}
P_{S_j}^{\mathcal{T}}=\exp\left(\frac{-N_0\beta_j}{\sigma_{S_jD_j}^2P_jr_{S_jD_j}^{-\alpha}}\right)\sum_{\substack{k\in\mathcal{T}\\k\ne j}}\left(\prod_{l\ne k}\frac{\theta_l\theta_k}{\theta_l-\theta_k}\right)\frac{1}{\left(\theta_k+1/\sigma_{S_jD_j}^2\right)}\label{12}
\end{equation*}
\begin{proof}
Let $\mathcal{A}_{\mathcal{T}^t}^t$ be the event that only nodes in set $\mathcal{T}^t$ of secondary nodes transmit in slot $t$ and let $\overline{\mathcal{O}_{S_jD_j,\mathcal{T}^t}^t}$ be the event of no outage on the $(S_j,D_j)$ link in slot $t$ when all nodes in the set $\mathcal{T}^t$ transmit.\\
The departure process of the $j$th secondary node can be written as:
\begin{equation}
Y_j^t=\sum_{\mathcal{T}^t\subseteq\{1,2,...,N\},j\in\mathcal{T}^t}\textbf{1}\left\{\left\{Q_P^t=0\right\}\bigcap \mathcal{A}_{\mathcal{T}_t}^t\bigcap\overline{\mathcal{O}_{S_jD_j,\mathcal{T}^t}^t}\right\}\label{10}
\end{equation}
By using the fact that if the primary queue is stable, then the process $\textbf{1}\{Q_P^t=0\}$ is stationary \cite{Szpankowski,Luo}; it can be easily shown that then, the process $Y_j^t$ is stationary. Hence, we drop the time indices. By Little's law \cite{bertsekas_book}, it follows that: $\mathrm{Pr}\left[Q_P=0\right]=1-\frac{\lambda_P}{\mu_P^{\mathrm{max}}}$.\\
Given a set $\mathcal{T}\subseteq\{1,2,...,N\}$ of secondary nodes transmitting in a slot, the probability that the secondary destination node $j\in \mathcal{T}$ is able to successfully decode the $j$th secondary source node transmission is given by:
\begin{equation}
P_{S_j}^{\mathcal{T}}=\mathrm{Pr}\left[\overline{\mathcal{O}_{S_jD_j,\mathcal{T}}}\right]=\mathrm{Pr}\left[\frac{P_{j}|h_{S_jD_j}|^2r_{S_jD_j}^{-\alpha}}{N_0+\sum_{k\in \mathcal{T},k\ne j}P_k|h_{S_kD_j}|^2r_{S_kD_j}^{-\alpha}}>\beta_{j}\right]\label{12}
\end{equation}
For the case of Rayleigh fading (Refer to Appendix A for the proof), $P_{S_j}^{\mathcal{T}}$ is given by:
\begin{equation}
P_{S_j}^{\mathcal{T}}=\exp\left(\frac{-N_0\beta_j}{\sigma_{S_jD_j}^2P_jr_{S_jD_j}^{-\alpha}}\right)\sum_{\substack{k\in\mathcal{T}\\k\ne j}}\left(\prod_{l\ne k}\frac{\theta_l\theta_k}{\theta_l-\theta_k}\right)\frac{1}{\left(\theta_k+1/\sigma_{S_jD_j}^2\right)}\label{7}
\end{equation}
where:\\
$\theta_k=\frac{P_jr_{S_jD_j}^{-\alpha}}{P_kr_{S_kD_j}^{-\alpha}\beta_j\sigma_{S_kD_j}^2}$.\\
By independence of the events in (\ref{10}), the throughput rate of the $j$th secondary source node is given by:
\begin{equation}
\lambda_j=\textbf{E}\left[Y_j^t\right]=\left(1-\frac{\lambda_P}{\mu_P^{\mathrm{max}}}\right)\sum_{\substack{\mathcal{T}\subseteq\{1,2,...,N\}\\j\in\mathcal{T}}}\prod_{k\in\mathcal{T}}q_k\prod_{l\in\mathcal{S}\setminus\mathcal{T}}(1-q_l)P_{S_j}^{\mathcal{T}}\label{13}
\end{equation}
Finally, the throughput region of the secondary network is obtained by taking the union over all possible transmission probability vectors $\textbf{q}=(q_1,q_2,...,q_N)\in[0,1]^N$ as in (\ref{9}).
\end{proof}

Next, we consider the symmetric case introduced in section(II). In that case, the probability of success of $j$th secondary node in the presence of $k$ other interfering transmissions is given by (Refer to Appendix A for the proof):
\begin{equation}
P^{(k)}=\mathrm{Pr}\left[\frac{P_0|h_{S_jD_j}|^2r_j^{-\alpha}}{N_0+\sum_{m=1}^k{P_0|\tilde{h_m}|^2r_j^{-\alpha}}}>\beta\right]=\mathrm{exp}\left(\frac{-\beta N_0}{\tilde{\sigma}^2 r_j^{-\alpha}P_0}\right)\frac{1}{\left(1+\beta\right)^k}\label{14}
\end{equation}
In this case, the throughput rate of the $j$th secondary node is given by:
\begin{align}
\lambda_j&=\left(1-\frac{\lambda_P}{\mu_P^{\mathrm{max}}}\right)q\sum_{k=0}^{N-1}\binom{N-1}{k}q^k(1-q)^{N-1-k}P^{(k)}\notag\\
&=\left(1-\frac{\lambda_P}{\mu_P^{\mathrm{max}}}\right)\mathrm{exp}\left(\frac{-\beta N_0}{\tilde{\sigma}^2 r_j^{-\alpha}P_0}\right)q \sum_{k=0}^{N-1}\binom{N-1}{k}q^k(1-q)^{N-1-k}\frac{1}{\left(1+\beta\right)^k}\notag\\
&=\left(1-\frac{\lambda_P}{\mu_P^{\mathrm{max}}}\right)\mathrm{exp}\left(\frac{-\beta N_0}{\tilde{\sigma}^2 r_j^{-\alpha}P_0}\right)q\left[1-q\frac{\beta}{1+\beta}\right]^{N-1}\label{15}
\end{align}
We note that due to perfect sensing, secondary nodes do not interfere with the primary. Hence, they can transmit at their maximum power in order to maximize their throughput rate $\left(\frac{\partial\lambda_j}{\partial P_0}>0\right)$ without affecting the stability of the primary queue and hence satisfying the cognitive principle of being transparent to the primary. This is not necessary true in the case of imperfect sensing as we will see in the next section.\\
Next, we calculate the optimum transmission probability $q$ at which secondary nodes should transmit in order to maximize their throughput. A very small $q$ will limit the interference between secondary nodes but will at the same time reduce the throughput while a large value of $q$ causes much interference between secondary nodes and hence also leads to a degradation in their throughput.
By setting $\frac{\partial\lambda_j}{\partial q}=0$, we get $q^*=\min\{1,\frac{1}{\alpha N}\}$, where $\alpha=\frac{\beta}{1+\beta}$. Thus, for small number of secondary nodes $N$, it is beneficial to transmit with probability one, while for a large value of $N$, secondary nodes should backoff to limit the interference on each other. Moreover, it is beneficial for both primary and secondary nodes that the primary node transmits at its maximum power and hence getting its maximum service rate and meanwhile maximizing the fraction $\left(1-\frac{\lambda_P}{\mu_P^{\max}}\right)$ of idle slots available to the secondary nodes. This may not be true if the sensing is not perfect because of the interference between primary and secondary nodes, so secondary nodes may suffer from degradation of their throughput if the primary node increases its transmission power.
\section{Imperfect Sensing Case}
Due to fading and other channel impairments, secondary nodes can encounter errors while sensing the channel and hence there is some possibility that they interfere with the primary node leading to a possible drastic reduction of its stable throughput. In this section, we quantify the effect of non ideal sensing on the throughput of primary and secondary nodes.
\subsection{Channel Sensing}
 Two errors may occur at the secondary nodes while sensing the channel, namely, false alarm and misdetection errors. All subsequent results are applicable for any sensing method as they are given in terms of general false alarm $P_f^{(i)}$ and misdetection $P_e^{(i)}$ probabilities at the $i$th secondary node. It should be noted that for a particular detector, $P_f^{(i)}$ and $P_e^{(i)}$ are related by its receiver operating characteristics (ROC)\cite{poor_book}.
\subsubsection{False Alarm Event}
False alarm occurs whenever the primary node is idle but is sensed to be busy. Clearly, false alarm errors do not affect the stable throughput of the primary but degrade the throughput of the secondary nodes.
\subsubsection{Misdetection Event}
It occurs when the primary node is busy but is sensed by some secondary nodes to be idle. Those secondary nodes will simultaneously transmit with the primary causing some interference at the primary destination. If the interference is strong enough, it may lead to instability of the primary queue.\\
Note that by the independence of the fading processes between nodes, the misdetection and false alarm events are independent between secondary nodes.
\subsection{Primary Queue Analysis}
\textit{Theorem 3:}\\
In the imperfect sensing case, the stability condition of the primary queue is given by:
\begin{equation}
\lambda_P<\mu_P=\sum_{\mathcal{U}\subseteq\{1,2,...,N\}}P_e^{\mathcal{U}}\left[\sum_{\mathcal{T}\subseteq\mathcal{U}}\left[\prod_{k\in\mathcal{T}}q_k\prod_{k\in\mathcal{U}\setminus\mathcal{T}}(1-q_k)\right]\mu_P^{(\mathcal{T})}\right]\label{20}
\end{equation}
where $P_e^{\mathcal{U}}$ and $\mu_P^{(\mathcal{T})}$ are given by (\ref{22}), (\ref{23}).
\begin{proof}
Let $\mathcal{U}^t\subseteq\mathcal{S}=\{1,2,...,N\}$ be the set (possibly empty) of secondary nodes that had misdetection probabilities at time slot $t$. A subset $\mathcal{T}^t\subseteq\mathcal{U}^t$ of these nodes will choose to transmit at that time slot.\\
The service process of the primary node can then be expressed as:
\begin{equation}
Y_P^t=\sum_{\mathcal{U}^t\subseteq\{1,2,...,N\}}\sum_{\mathcal{T}^t\subseteq\mathcal{U}^t}\textbf{1}\left\{\mathcal{E}_{\mathcal{U}^t}^t\bigcap \mathcal{A}_{\mathcal{T}^t}^t\bigcap\overline{\mathcal{O}_{S_PD_P,\mathcal{T}^t}^t}\right\}\label{21}
\end{equation}
where $\mathcal{E}_{\mathcal{U}^t}^t$ denotes the event that only nodes in the set $\mathcal{U}^t$ commit an error in detection of the primary node in time slot $t$; $\mathcal{A}_{\mathcal{T}^t}^t$ is the event that only nodes in the set $\mathcal{T}^t$ transmit at time slot $t$ and $\overline{\mathcal{O}_{S_PD_P,\mathcal{T}^t}^t}$ is the event of no outage on the $(S_P,D_P)$ link in the presence of an interfering set $\mathcal{T}^t$ of secondary nodes.
The process $Y_P^t$ is clearly stationary, thus we drop the time indices $t$ subsequently.\\
By independence, the probability that only nodes in the set $\mathcal{U}$ have misdetection is given by:
\begin{equation}
P_e^{\mathcal{U}}=\mathrm{Pr}\left[\mathcal{E_U}\right]=\prod_{j\in\mathcal{U}}P_e^{(j)}\prod_{j\in\mathcal{S}\setminus\mathcal{U}}\left(1-P_e^{(j)}\right)\label{22}
\end{equation}
The service rate at the primary node given a set $\mathcal{T}$ of transmitting nodes -as defined above- is given by:
\begin{equation}
\mathrm{Pr}\left[\overline{\mathcal{O}_{S_PD_P,\mathcal{T}}}\right]=\mu_P^{(\mathcal{T})}=\mathrm{Pr}\left[\frac{P_P|h_{S_PD_P}|^2r_{S_PD_P}^{-\alpha}}{N_0+\sum_{k\in\mathcal{T}}P_{S_k}|h_{S_kD_P}|^2r_{S_kD_P}^{-\alpha}}>\beta_P\right]\label{23}
\end{equation}
Noting that $\mu_P^{(\varnothing)}=\mu_P^{\mathrm{max}}$, we can get the average service rate at the primary queue as given in (\ref{23}), and hence by Loynes' theorem, the proof is complete.
\end{proof}
Note that from equation (\ref{23}), $\mu_P^{(\mathcal{T})}<\mu_P^{\max}$ unless $\mathcal{T}$ is the empty set.\\ Using that $\sum_{\mathcal{U}\subseteq\{1,2,...,N\}}P_e^{\mathcal{U}}\left[\sum_{\mathcal{T}\subseteq\mathcal{U}}\left[\prod_{k\in\mathcal{T}}q_k\prod_{k\in\mathcal{U}\setminus\mathcal{T}}(1-q_k)\right]\right]=1$, we get that equation (\ref{20}) is a convex combination of terms less than or equal to $\mu_P^{\max}$, hence $\mu_P$ given in equation (\ref{20}) is strictly less than $\mu_P^{\max}$, which is an expected result due to secondary interference.\\
Next, we specialize to the symmetric case with Rayleigh fading. In this case, by symmetry, the probability of misdetection is the same for all the secondary nodes,i.e., $P_e^{(j)}=P_e,\mbox{ }j\in\{1,2,...,N\}$.

Let $\mu_P^{(k)}$ be the success probability of primary node given $(k)$ secondary concurrent transmissions, then by similar analysis as in Appendix A, we get $\mu_P^{(k)}$ as:
\begin{align}
\mu_P^{(k)}&=\mathrm{Pr}\left[\frac{P_P|h_{S_pD_P}|^2r_{S_PD_P}^{-\alpha}}{N_0+\sum_{j=1}^k{P_0|h_j|^2r_0^{-\alpha}}}>\beta_P\right]
=\mathrm{exp}\left(\frac{-N_0\beta_P}{\sigma_{S_pD_P}^2P_Pr_{S_PD_P}^{-\alpha}}\right)\frac{1}{{\left(1+\frac{P_0r_0^{-\alpha}\beta_P\sigma_0^2}{\sigma_{S_PD_P}^2P_Pr_{S_pD_P}^{-\alpha}}\right)}^k}\notag\\
&=\mu_P^{\mathrm{max}}{\left(\frac{a}{1+a}\right)}^k\label{25}
\end{align}
where:
\begin{equation}
a=\frac{\sigma_{S_PD_P}^2P_Pr_{S_PD_P}^{-\alpha}}{\sigma_0^2\beta_PP_0r_0^{-\alpha}}\label{26}
\end{equation}
By symmetry, the average service rate of the primary queue is given by:
\begin{align}
\mu_P&=\sum_{L=0}^N\binom{N}{L}P_e^L(1-P_e)^{N-L}\left[\sum_{k=0}^L\binom{L}{k}q^k(1-q)^{L-k}\mu_P^{(k)}\right]=\mu_P^{\mathrm{max}}\left[1-\frac{qP_e}{a+1}\right]^N\notag\\
&=\mu_P^{\mathrm{max}}\left[1-qP_e\frac{P_0r_0^{-\alpha}\beta_P\sigma_0^2}{\sigma_{S_PD_P}^2P_Pr_{S_PD_P}^{-\alpha}+\sigma_0^2P_0r_0^{-\alpha}\beta_P}\right]^N\label{27}
\end{align}
The effect of imperfect sensing is shown in the multiplication of $\mu_P^{\mathrm{max}}$ by a term less than one.\\
For stability, and applying Loynes' theorem, we should have:
\begin{equation}
\lambda_P<\mu_P^{\mathrm{max}}\left[1-qP_e\frac{P_0r_0^{-\alpha}\beta_P\sigma_0^2}{\sigma_{S_PD_P}^2P_Pr_{S_PD_P}^{-\alpha}+\sigma_0^2P_0r_0^{-\alpha}\beta_P}\right]^N\label{28}
\end{equation}

The primary user will choose its arrival rate $\lambda_P<\mu_P^{\max}$ independently of the secondary network. In the imperfect sensing case, $\mu_P<\mu_P^{\max}$, and hence, the secondary nodes should limit their transmission power and/or transmission probabilities to limit the interference on the primary node and hence ensuring that its arrival rate $\lambda_P$ be less than $\mu_P$ to avoid the instability of its queue.

It is straightforward to establish the following properties about $\mu_P$ given by (\ref{27}).\\
\textit{Proposition 1:}\\
The primary node service rate in the non-ideal sensing case, as given by (\ref{27}) satisfies:
\begin{equation*}
\mbox{(i) }0\leq\mu_P\leq\mu_P^{\mathrm{max}}
\end{equation*}
\begin{equation*}
\mbox{(ii) }\lim_{a\to\infty}\mu_P=\mu_P^{\mathrm{max}}
\end{equation*}
\begin{equation*}
\mbox{(iii) }
    \text{$\lim_{a\to0}\mu_P$} =
      \begin{cases}
       \mu_P^{\mathrm{max}}\left[1-qP_e\right]^N \text{if }P_P>0\text{ and }P_0\to\infty \\
        0\text{ if }P_P\to0.
      \end{cases}
     \end{equation*}
\begin{equation*}
\mbox{(iv) }\lim_{q\to0}\mu_P=\mu_P^{\mathrm{max}}
\end{equation*}
\begin{equation*}
\mbox{(v) }\lim_{q\to1}\mu_P=\mu_P^{\mathrm{max}}\left[1-\frac{P_e}{a+1}\right]^N
\end{equation*}
\begin{equation*}
\mbox{(vi) }\frac{d}{da}\mu_P>0,\mbox{ i.e., }\mu_P\mbox{ is strictly increasing with }a.
\end{equation*}
\begin{equation*}
\mbox{(vii) }\frac{d}{dq}\mu_P<0,\mbox{ i.e., }\mu_P\mbox{ is strictly decreasing with }q.
\end{equation*}
From proposition $1$, we can draw some important conclusions: property (i) states that the effect of non ideal sensing at the secondary nodes is the degradation of the service rate of the primary licensed node due to the interference from secondary nodes on the primary. Such negative effect does not occur in the perfect sensing case where the service rate of the primary node is independent of secondary transmissions. Properties (ii),(iii) and (vi) reveal that unless $a\to\infty$ i.e. either $P_P\to\infty$ or $P_0\to0$, primary node cannot achieve its maximum service rate $\mu_P^{\mathrm{max}}$ achieved in the case of perfect sensing. Also, for fixed $P_P$, which is the case of interest, secondary nodes have a maximum power, possibly infinite if $\lambda_P$ is small enough, at which they can transmit without affecting the stability of the primary node. Moreover, even if the interference of the secondary nodes is very high (case of $P_0\to\infty$), due to the random access of the secondary nodes to the channel, primary node can still achieve a portion of its maximum service rate $\mu_P^{\mathrm{max}}$ because there is a positive probability that none of the secondary nodes will transmit in a given slot. Finally, properties (iv), (v) and (vii) suggest that for fixed $P_P$ and $P_0$, secondary nodes can control their interference level on the primary by adjusting their transmission probabilities $q$ which is sometimes easier to implement than power control due to hardware complexity and non linearity of the power amplifiers needed for power control over wide range.\\
For $\lambda_P<\mu_P$ to be satisfied, we can solve for minimum value of $a=\frac{\sigma_{S_PD_P}^2P_Pr_{S_PD_P}^{-\alpha}}{\sigma_0^2\beta_PP_0r_0^{-\alpha}}$ and for the maximum value of $q$ to get the maximum possible transmission power $(P_0)$ and maximum possible transmission probability $(q)$ of the secondary nodes while remaining ``transparent'' to the primary, i.e., without affecting its stability.\\
By using equation (\ref{27}) and proposition $1$, we obtain:
\begin{equation}
    \text{$a_{\mathrm{min}}$} =
      \begin{cases}
       \frac{qP_e}{1-(\lambda_P/\mu_P^{\mathrm{max}})^{1/N}}-1 \hspace{.7cm}\text{ if }\mu_P^{\mathrm{max}}\left(1-qP_e\right)^N<\lambda_P<\mu_P^{\max}\\
        0\hspace{.7cm}\text{ otherwise }.
      \end{cases}\label{29}
\end{equation}
Hence:
\begin{equation}
    \text{$q<q_{\mathrm{max}}$} =
      \begin{cases}
       1 \hspace{1cm}\mbox{ if }\lambda_P<\mu_P^{\mathrm{max}}\left[1-\frac{P_e}{a+1}\right]^N \\
        \left[1-\left(\frac{\lambda_P}{\mu_P^{\mathrm{max}}}\right)^{1/N}\right]\frac{a+1}{P_e}\\\hspace{1.1cm}\mbox{ if }\mu_P^{\mathrm{max}}\left[1-\frac{P_e}{a+1}\right]^N<\lambda_P<\mu_P^{\max}
      \end{cases}\label{30}
\end{equation}
For fixed primary transmission power $P_P$, we can calculate the maximum transmission power allowed at secondary nodes as:
\begin{equation}
    \text{$P_0$} < P_0^{\max}=
      \begin{cases}
      \infty\hspace{.7cm}\text{ if }\lambda_P<\mu_P^{\mathrm{max}}\left(1-qP_e\right)^N\\
       \frac{\sigma_{S_PD_P}^2P_Pr_{S_PD_P}^{-\alpha}}{r_0^{-\alpha}\beta_P\sigma_0^2}\left[\frac{1-(\lambda_P/\mu_P^{\max})^{1/N}}{qP_e-1+(\lambda_P/\mu_P^{\max})^{1/N}}\right]\\ \hspace{1cm}\mbox{ if }\mu_P^{\max}\left(1-qP_e\right)^N<\lambda_P<\mu_P^{\max}
      \end{cases}\label{31}
\end{equation}
From equations (\ref{30}) and (\ref{31}), we conclude that for fixed $P_P$, if $\lambda_P<\mu_P^{\mathrm{max}}\left[1-\frac{P_e}{a+1}\right]^N$, secondary nodes can transmit at any desired chosen probability without affecting the stability of the primary while they have to backoff to reduce their interference on the primary node if $\mu_P^{\mathrm{max}}\left[1-\frac{P_e}{a+1}\right]^N<\lambda_P<\mu_P^{\max}$. On the other hand, for fixed transmission probability $q$, if $\lambda_P<\mu_P^{\max}(1-qP_e)^N$, secondary nodes can transmit at any power without affecting the stability of the primary node while there exists a finite maximum allowed power if $\lambda_P>\mu_P^{\max}(1-qP_e)^N$. This can be understood by noting that $(1-qP_e)^N$ is the probability that none of the secondary source nodes transmit in a slot while the primary is busy, attracting the attention to the benefit of using random access as a multiple access protocol in the secondary network. Note that, in practical situations, the transmit power of a node is also limited by the power amplifier used, but we ignore this aspect here.
\subsection{Secondary Queues}
A secondary node gets a packet served in the imperfect sensing case, either if the primary node is idle with no false alarm occurring at that secondary node and that node transmits and is successful, or if the primary node is busy with an incorrect detection of the primary node occurring at that secondary node and the secondary node transmits and is successful.\\
\textit{Theorem 4:}\\
The throughput of the $j$th secondary source node in the imperfect sensing case is given by:
\begin{equation}
\lambda_{j}=\left(1-\frac{\lambda_P}{\mu_P}\right)\lambda_{j}^{\text{(P idle)}}+\left(\frac{\lambda_P}{\mu_P}\right)\lambda_{j}^{\text{(P busy)}}\label{32}
\end{equation}
where:\\
$\lambda_{j}^{\text{(P idle)}}$ and $\lambda_{j}^{\text{(P busy)}}$ are the average throughput rates of the $jth$ secondary node given that the primary node is idle, and  busy respectively. and are given by (\ref{38}), (\ref{39}).
\begin{proof}
By the saturation assumption of the secondary queues, the average service rate of the primary node is independent of the secondary nodes' queue states (i.e. there is no queueing interactions) and hence, the probability that the primary node is idle = 1- Probability that primary node is busy = $1-\frac{\lambda_P}{\mu_P}$.\\
The departure process at the $j$th secondary source node is given by:
\begin{align}
Y_j^t=&\sum_{\mathcal{F}^t\subseteq\{\mathcal{S}\setminus\{j\}\}}\;\sum_{\mathcal{T}^t\subseteq\{\mathcal{S}\setminus\mathcal{F}^t\},\,j\in\mathcal{T}^t}\textbf{1}\left\{\{Q_P^t=0\}\bigcap\mathcal{F}^t\bigcap \mathcal{A}_{\mathcal{T}^t}^t\bigcap\overline{\mathcal{O}_{S_jD_j,\mathcal{F}^t,\mathcal{T}^t}^t}\right\}+\notag\\
&\sum_{\mathcal{E}^t\subseteq\mathcal{S},\,j\in\mathcal{E}^t}\;\sum_{\mathcal{T}^t\subseteq\mathcal{E}^t,\,j\in\mathcal{T}^t}\textbf{1}\left\{\{Q_P^t\ne0\}\bigcap\mathcal{E}^t\bigcap \mathcal{A}_{\mathcal{T}^t}^t\bigcap\overline{\mathcal{O}_{S_jD_j,\mathcal{E}^t,\mathcal{T}^t}^t}\right\}\label{33}
\end{align}
where $\mathcal{S}=\{1,2,...,N\}$, $\mathcal{F}^t$ is the event that only the nodes in set $\mathcal{F}$ have a false alarm in slot $t$ whenever the primary source node is idle, $\mathcal{E}^t$ is the event that only the nodes in set $\mathcal{E}$ have a misdetection of the primary node in slot $t$ whenever the primary is busy, $\mathcal{A}_{\mathcal{T}^t}^t$ is the event that only nodes in set $\mathcal{T}^t$ transmit at time slot $t$. The event $\overline{\mathcal{O}_{S_jD_j,\mathcal{F}^t,\mathcal{T}^t}^t}$ is the event of no outage on the $j$th secondary source-destination link when the set $\mathcal{F}$ of nodes has false alarm and nodes in the set $\mathcal{T}$ of secondary nodes transmit simultaneously at time slot $t$, while the event $\overline{\mathcal{O}_{S_jD_j,\mathcal{E}^t,\mathcal{T}^t}^t}$ is the event of no outage on the $j$th secondary source-destination link when the set $\mathcal{E}$ of nodes has misdetection of the activity of the primary node, and nodes in the set $\mathcal{T}$ of secondary nodes transmit simultaneously.\\
Since the process is stationary, we can drop the time indices.\\
The $jth$ secondary node departure rate can be written as:
\begin{equation}
\lambda_{j}=\textbf{E}[Y_j^t]=\left(1-\frac{\lambda_P}{\mu_P}\right)\lambda_{j}^{\text{(P idle)}}+\left(\frac{\lambda_P}{\mu_P}\right)\lambda_{j}^{\text{(P busy)}}\label{34}
\end{equation}
If the primary node is idle, then the probability that a set $\mathcal{F}\subseteq\{\mathcal{S}\setminus\{j\}\}$ of secondary nodes has false alarms while all other secondary nodes do not is:
\begin{equation}
\text{Pr[$\mathcal{F}$ $|$ P is idle]}=\prod_{i\in \mathcal{F}}P_f^{(i)}\prod_{i\in\mathcal{S}\setminus \mathcal{F}}\left(1-P_f^{(i)}\right)\label{35}
\end{equation}
then we can write:
\begin{equation}
\lambda_{j}^{\text{(P idle)}}=\sum_{\mathcal{F}\subseteq\mathcal{S}\setminus\{j\}}\text{ Pr[$\mathcal{F}$ $|$ P is idle] }\lambda_{j}^{\text{(P idle, $\mathcal{F}$)}}\label{36}
\end{equation}
where:
\begin{equation}
\lambda_{j}^{\text{(P idle, $\mathcal{F}$)}}=\sum_{\mathcal{T}\subseteq\mathcal{S}\setminus \mathcal{F},\,j\in\mathcal{T}}\left[\prod_{i\in\mathcal{T}}q_i\prod_{i\in\mathcal{T}^c}(1-q_i)\right]\mathrm{Pr}\left[\frac{P_{j}|h_{S_jD_j}|^2r_{S_jD_j}^{-\alpha}}{N_0+\sum_{l\in\mathcal{T}}P_{l}|h_{S_lD_j}|^2r_{S_lD_j}^{-\alpha}}>\beta_{j}\right]\label{37}
\end{equation}
where $\mathcal{T}^c=\mathcal{S}\setminus \{\mathcal{F}\bigcup\mathcal{T}\}$. Hence,
\begin{equation}
\lambda_{j}^{\text{(P idle)}}=\sum_{\mathcal{F}\subseteq\mathcal{S}\setminus\{j\}}\left[\prod_{i\in \mathcal{F}}P_f^{(i)}\prod_{i\in\mathcal{S}\setminus \mathcal{F}}\left(1-P_f^{(i)}\right)\right]\lambda_{j}^{\text{(P idle, $\mathcal{F}$)}}\label{38}
\end{equation}
Similarly:
\begin{equation}
\lambda_{j}^{\text{(P busy)}}=\sum_{\mathcal{E}\subseteq\mathcal{S},\,j\in\mathcal{E}}\left[\prod_{i\in \mathcal{E}}P_e^{(i)}\prod_{i\in\mathcal{S}\setminus{\mathcal{E}}}\left(1-P_e^{(i)}\right)\right]\lambda_{j}^{\text{(P busy, $\mathcal{E}$)}}\label{39}
\end{equation}
where:
\begin{equation}
\lambda_{j}^{\text{(P busy, $\mathcal{E}$)}}=\sum_{\mathcal{T}\subseteq \mathcal{E},\,j\in\mathcal{T}}\left[\prod_{i\in\mathcal{T}}q_i\prod_{i\in \mathcal{E}\setminus{\mathcal{T}}}(1-q_i)\right]\mathrm{Pr}\left[\frac{P_{j}|h_{S_jD_j}|^2r_{S_jD_j}^{-\alpha}}{N_0+P_P|h_{S_PD_j}|^2r_{S_PD_j}^{-\alpha}+\sum_{l\in\mathcal{T}}P_{l}|h_{S_lD_j}|^2r_{S_lD_j}^{-\alpha}}>\beta_{j}\right]\label{40}
\end{equation}
Finally, the throughput region of the secondary nodes is given by:
\begin{equation}
\mathcal{L}=\bigcup_{(q_1,q_2,...,q_N)\in[0,1]^N}\Big\{(\lambda_1,\lambda_2,...,\lambda_N)\Big\}\label{41}
\end{equation}
where $\lambda_j$ is given by (\ref{34}) for $j\in\{1,2,...,N\}$.
\end{proof}
Next, we specialize to the case of symmetric secondary cluster introduced in section (II).
In this case, $P_e^{(j)}=P_e$ and $P_f^{(j)}=P_f$ for all $j\in\{1,2,...,N\}$. The average throughput at the $j$th secondary node can be written as:
\begin{align}
\lambda_j=&\left(1-\frac{\lambda_P}{\mu_P}\right)q(1-P_f)\sum_{L=0}^{N-1}\binom{N-1}{l}(1-P_f)^LP_f^{N-1-L}\left[\sum_{k=0}^L\binom{L}{k}q^k(1-q)^{L-k}\lambda_j^{\text{(idle,k)}}\right]+\notag\\
&\left(\frac{\lambda_P}{\mu_P}\right)qP_e\sum_{L=0}^{N-1}\binom{N-1}{L}P_e^L(1-P_e)^{N-1-L}\left[\sum_{k=0}^L\binom{L}{k}q^k(1-q)^{L-k}\lambda_j^{\text{(busy,k)}}\right]\label{42}
\end{align}
where:
\begin{equation}
\lambda_j^{\text{(idle,k)}}=\mathrm{Pr}\left[\frac{P_0|h_{S_jD_j}|^2r_j^{-\alpha}}{N_0+\sum_{l=1}^kP_0|\tilde{h_l}|^2r_j^{-\alpha}}>\beta\right]=\mathrm{exp}\left(\frac{-\beta N_0}{\tilde{\sigma}^2P_0r_j^{-\alpha}}\right)\frac{1}{(1+\beta)^k}\label{43}
\end{equation}
\begin{align}
\lambda_j^{\text{(busy,k)}}&=\mathrm{Pr}\left[\frac{P_0|h_{S_jD_j}|^2r_j^{-\alpha}}{N_0+\sum_{l=1}^kP_0|\tilde{h_l}|^2r_j^{-\alpha}+P_P|h_{S_pD_j}|^2r_j^{-\alpha}}>\beta\right]\notag\\
&=\mathrm{exp}\left(\frac{-\beta N_0}{\tilde{\sigma}^2P_0r_j^{-\alpha}}\right)\left(1+\frac{P_Pr_{S_PD_j}^{-\alpha}\beta\sigma_{S_PD_j}^2}{P_0r_j^{-\alpha}\tilde{\sigma}^2}\right)^{-1}\frac{1}{(1+\beta)^k}\label{44}
\end{align}
After some algebra, the throughput of the $j$th secondary source node can be written as:
\begin{align}
\lambda_j=&\left(1-\frac{\lambda_P}{\mu_P}\right)\mathrm{exp}\left(\frac{-\beta N_0}{\tilde{\sigma}^2P_0r_j^{-\alpha}}\right)q(1-P_f)\left[1-q(1-P_f)\frac{\beta}{\beta+1}\right]^{N-1}+\notag\\
&\left(\frac{\lambda_P}{\mu_P}\right)\mathrm{exp}\left(\frac{-\beta N_0}{\tilde{\sigma}^2P_0r_j^{-\alpha}}\right)qP_e\left(1+\frac{P_Pr_{S_PD_j}^{-\alpha}\beta\sigma_{S_PD_j}^2}{P_0r_j^{-\alpha}\tilde{\sigma}^2}\right)^{-1}\left[1-qP_e\frac{\beta}{\beta+1}\right]^{N-1}\label{45}
\end{align}
Secondary nodes will try to maximize their throughputs (for example sum throughput) by optimizing $P_0$ and $q$ subject to the constraints given by (\ref{30}) and (\ref{31}). That is by solving:
\begin{align*}
&\max_{P_0\geq 0,\mbox{ } q\geq 0}\sum_{j=1}^N\lambda_j\notag\\
&\mbox{s.t.  (\ref{30}) and (\ref{31})}\notag\\
&\mbox{where }\lambda_j\mbox{ is given by (\ref{45})}\label{90}
\end{align*}
The objective function is however non-convex and the solution is hard to find in closed form. In section VII, we present numerical solution to that optimization problem and in the following, we present some intuitive properties of the solution based on the structure of the objective function.
Clearly, the smaller the value of the false alarm probability $P_f$, the higher the fraction of idle slots in which the secondary nodes access the channel without any interference from the primary. The secondary throughput of node $j$ $(\lambda_j)$ decreases with $P_f$ because of the smaller fraction of idle slots accessed by the secondary nodes but also increases with $P_f$ due to less interference between secondary nodes. Such variation depends on other parameter values, but in general, if $N$ is small, then the first effect can be significant while if $N$ is large enough, then only the second effect becomes prevalent. Let $I=\frac{P_Pr_{S_PD_j}^{-\alpha}\sigma_{S_PD_j}^2\beta}{P_0r_j^{-\alpha}\tilde{\sigma}^2}$, which is proportional to the ratio of the interference of the primary on the secondary to the secondary's power. If this term is large, then the primary highly interferes with the secondary and the throughputs of the secondary nodes when the primary is busy are largely reduced. In this case, the secondary throughput is dominated by the first term of (\ref{45}). On the other hand, if $I$ is small enough, then the second term of (\ref{45}) might become significant and the interference from the primary does not significantly reduce the throughputs of the secondary nodes. In this case, $\lambda_j$ increases with increasing $P_e$ because of more opportunities for secondary nodes to transmit when the primary is busy despite the fact that the term $[1-qP_e\frac{\beta}{\beta+1}]^{N-1}$ decreases with $P_e$ because of the secondary interference on each other. The variation of $\lambda_j$ with $\frac{\lambda_P}{\mu_P^{\max}}$ is more subtle. One one hand, the fraction of idle slots $1-\frac{\lambda_P}{\mu_P}$ decreases with increasing $\frac{\lambda_P}{\mu_P^{\max}}$; but also $\frac{\lambda_P}{\mu_P}$ which represents the fraction of busy slots increases. Such variation is highly dependable on the other parameters. For small values of $\frac{\lambda_P}{\mu_P^{\max}}$, secondary nodes (by (\ref{31})) can transmit at their maximum power and get an increasing throughput due to increasing of the fraction of busy slots especially when $P_e$ is high until $\frac{\lambda_P}{\mu_P^{\max}}$ reaches a value at which transmission power $P_0$ and/or transmission probabilities $q$ should decrease to limit the interference on the primary and thus secondary throughput decrease as well. Finally, it should be noted that sensing errors might lead to higher secondary throughput compared with the perfect sensing case (in contrast with the primary node where imperfect sensing always leads to lower maximum stable throughput). This can be explained by observing that incorrect sensing gives the secondary nodes more opportunities for transmission during the busy slots of the primary which might lead to a net increase in throughput especially when $\frac{\lambda_P}{\mu_P^{\max}}$ is large. Such observations will become clearer in section VII.
\section{Relaying in the Perfect Sensing Case}
Primary users would be willing to share their channel resources with secondary users if such sharing will benefit them. Hence, forcing the secondary nodes to relay the primary node's unsuccessful packets would be the price the secondary nodes pay for using the channels and the incentive to the primary users to share the channel. Moreover, by relaying the primary packets, secondary nodes might benefit by increasing the number of idle slots available for secondary transmissions. In this section, we propose and analyze a simple distributed cooperative protocol between the secondary and primary nodes. We restrict the analysis to the perfect sensing case but the imperfect sensing case can be handled similarly. We also restrict the analysis to the channel symmetric case as described previously. Similar results can be shown for the asymmetric case with any fading distribution. Refer to Appendix (B) for the proof in the general case.
\subsection{Relaying Protocol}
The relaying protocol achieves throughput gain with no channel state information (CSI) about $h_{S_kD_P}$ fading coefficients available at the secondary nodes by using Distributed Orthogonal Space-Time Block Code (D-OSTBC). That is, each of the secondary nodes that are able to successfully decode a primary's unsuccessful packet will mimic an antenna in a regular Space-Time Code (STC) setting of a multiple-input single-output (MISO) channel. Such OSTBC always exists for one dimensional signal constellations for any number of relaying nodes \cite{Tarokh_2}. In this case, these OSTBC schemes achieve full diversity gain and/or power gain at coding rate =1 while ensuring simple decoding rule based on linear processing at the receiver.\\
\textbf{\textit{Remarks:}}\\
1. Each of the relaying nodes must know which antenna it mimics in the underlying STC used, which can be achieved by either some coordination between the secondary nodes or by a prior node indexing and observing ACK/NACK packets generated by the secondary nodes regarding the primary packet. Those ACK/NACK messages are assumed to be available to all nodes throughout the network.\\
2. If the packet length is not an integer multiple of the number of relaying nodes, the last block of symbols in the packet is relayed by a smaller number of nodes. However, such effect is typically small as packet lengths are much longer than the number of relaying nodes and thus, we ignore such ``edge effects'' in the sequel.\\
3. For two-dimensional constellations, it is shown in \cite{ST_3} that the rates of complex orthogonal space–time block codes for more than two transmit antennas are upper-bounded by 3/4 while the rates of generalized complex orthogonal space–time block codes for more than two transmit antennas are upper-bounded by 4/5. In this paper, we mainly focus on the scenario where the rate is fixed in order to avoid both analytical and practical issues (such as synchronization problems) related to variable rate systems.\\
The relaying protocol works as follows:
At every busy slot of the primary node, if one or more secondary nodes are able to successfully decode the packet sent by primary node while the primary destination can not, then these secondary nodes store this packet in a special queue (relaying queue) and then send an ACK feedback to the primary and the primary node releases the packet from its queue. We assume that this ACK messages will also be heard by all the secondary nodes and thus the secondary nodes which could not receive that packet will abstain from transmission until that packet is successfully delivered to the primary destination and thus avoid interfering with the primary relayed packets. In the next available primary user's idle slot, the secondary nodes which were able to decode the primary packet will transmit it using D-OSTBC as described above. It should be noted that the primary packets are given priority for transmission, i.e., a secondary source node will not transmit its own packets unless it does not have any primary packets to relay and none of the other secondary nodes has any.
\subsection{Protocol Analysis}
We proceed by proving the main result for symmetric network and for arbitrary fading distributions and then we will provide closed form expressions for the special case of Rayleigh fading. For the case of asymmetric network, similar results hold and the proof is in Appendix (B).\\
For a secondary source node to successfully decode a primary packet, the minimum SNR value needed is $\beta_P$.
Let $P_d$ be the probability that one of the secondary nodes be able to successfully decode the primary packet (which is same for all secondary nodes by symmetry), then:
\begin{equation}
P_d=\mathrm{Pr}\left[\frac{P_P|h_{S_PS_j}|^2r_{S_PS_j}^{-\alpha}}{N_0}>\beta_P\right]=\mathrm{Pr}\left[|h_{S_PS_j}|^2>\frac{\beta_PN_0}{P_Pr^{-\alpha}}\right]\label{46}
\end{equation}
Let $M^t$ be a random variable denoting the number of secondary source nodes that successfully decoded a primary packet in time slot $t$, then:
\begin{equation}
\mathrm{Pr}[M^t=m]=\binom{N}{m}P_d^m(1-P_d)^{N-m},\mbox{    }m=0,1,...,N\label{47}
\end{equation}
According to the above described protocol, the primary queue service process takes the form:
$Y_P^t=\textbf{1}\left\{\overline{\mathcal{O}_{S_PD_P}^t}\bigcup\overline{\mathcal{O}_{S_PS_S}^t}\right\}$,
where $\overline{\mathcal{O}_{S_PD_P}^t}$ and $\overline{\mathcal{O}_{S_PS_S}^t}$ denote the events of no outage on the primary source-primary destination link and the event that at least one secondary source node was able to successfully decode the packet respectively. Clearly, the service process at the primary source queue is stationary. Hence, the success probability is given by:
\begin{equation}
\mu_P=\textbf{E}\left[Y_P^t\right]=\mu_P^{\mathrm{max}}+[1-(1-P_d)^N]-\mu_P^{\mathrm{max}}[1-(1-P_d)^N]=1-(1-\mu_P^{\max})(1-P_d)^N\label{49}
\end{equation}
which is strictly greater than $\mu_P^{\max}$.\\
It should be noted that stability of the relaying queues at the secondary nodes (as will come clearer later, see equation (\ref{56})) guarantees that the primary's packets that were successfully decoded at the secondary nodes will be eventually delivered to the primary destination because the secondary nodes' relaying queues -due to their stability- will empty infinitely often.\\
\textit{Theorem 5:}\\
Under the previously described relaying protocol, the stability condition of the system is:
\begin{equation}
\lambda_P<\frac{\mu_PP_s}{P_s+(1-\mu_P^{\max})P_d}\label{50}
\end{equation}
where:
\begin{equation}
P_s=\sum_{k=0}^{N-1}\binom{N-1}{k}P_d^k(1-P_d)^{N-1-k}\mathrm{Pr}\left[\sum_{i=1}^{k+1}|h_i|^2>\frac{\beta_PN_0}{P_0r_0^{-\alpha}}\right]\label{51}
\end{equation}
\begin{proof}
Each secondary source node will have exogenous packet arrivals from the primary source node to relay in the subsequent idle slots in addition to the secondary packets to be transmitted. The arrival process to the secondary source node from the primary is given by:
\begin{equation}
X_{ext}^P=\textbf{1}\left\{\{Q_P^t\ne0\}\bigcap\mathcal{O}_{S_PD_P}^t\bigcap\overline{\mathcal{O}_{S_PS}^t}\right\}\label{52}
\end{equation}
where $\mathcal{O}_{S_PD_P}^t$ is the outage event on the primary source-destination link and $\overline{\mathcal{O}_{S_PS}^t}$ is the event of no outage on the primary source-secondary source link at time slot $t$.\\
The SNR per symbol of the relayed packet at the primary destination node given that $k$ secondary nodes transmit is given by:
\begin{equation}
SNR=\frac{P_0r_0^{-\alpha}\sum_{i=1}^k|h_i|^2}{N_0}\label{53}
\end{equation}
Hence, the probability of no outage given $k+1$ nodes transmit in slot $t$ is given by:
\begin{equation}
\mathrm{Pr}[\overline{\mathcal{O}_{SD_P,k+1}^t}]=\mathrm{Pr}\left[\sum_{i=1}^{k+1}|h_i|^2>\frac{\beta_PN_0}{P_0r_0^{-\alpha}}\right]\label{54}
\end{equation}
The service process of the primary packets queued at a secondary source node is given by:
\begin{equation}
Y_{ext}^P=\sum_{k=0}^{N-1}\textbf{1}\left\{\{Q_P^t=0\}\bigcap\{\bar{M}^t=k\}\bigcap\overline{\mathcal{O}_{SD_P,k+1}^t}\right\}\label{55}
\end{equation}
where $\bar{M}$ is a random variable denoting the number of other secondary nodes that could decode the packet in service, and $\overline{\mathcal{O}_{SD_P,k+1}^t}$ is the event of no outage at the primary destination when $k+1$ secondary nodes collaboratively transmit the relayed primary packet.\\
The arrival and service processes $X_{ext}^P$ and $Y_{ext}^P$ are jointly stationary and hence by Lemma 1, we can get the condition of stability as:
\begin{equation}
\lambda_{ext}^P=\textbf{E}\left[X_{ext}^P\right]=\left(\frac{\lambda_P}{\mu_P}\right)(1-\mu_P^{\max})P_d<\mu_{ext}^P=\textbf{E}\left[Y_{ext}^P\right]=\left(1-\frac{\lambda_P}{\mu_P}\right)P_s\label{56}
\end{equation}
where $P_s$ is as given by (\ref{51}).\\
For stability of the primary queue, we should also have that $\lambda_P<\mu_P$ and (\ref{56}) satisfied, leading to:
\begin{equation}
\lambda_P<\lambda_P^{\max}=\frac{\mu_PP_s}{P_s+(1-\mu_P^{\max})P_d}\label{57}
\end{equation}
where $\lambda_P^{\max}$ is the maximum stable throughput rate of the primary queue.
\end{proof}
\textit{Proposition 2:}\\
The success probability $P_s$ as given by (\ref{51}) is strictly increasing with $N$. Moreover, as $N\to\infty$, $P_s\to 1$.\\
\textit{Proof:} Refer to Appendix C.

\textit{Proposition 3:}\\
The maximum possible arrival rate at the primary node that keeps the system stable as given by (\ref{57}) is higher than in the case of no-relaying only if $\mu_P^{\max}<\frac{P_{\mathrm{s}}[1-(1-P_d)^N]}{P_d}$.
\begin{proof}
Follows immediately by setting:
\begin{equation*}
\mu_P^{\max}<\frac{\mu_PP_{\mathrm{s}}}{P_{\mathrm{s}}+(1-\mu_P^{\max})P_d}\qedhere
\end{equation*}
and substituting $\mu_P$ from (\ref{49}).
\end{proof}
The term $\frac{1-(1-P_d)^N}{P_d}$ is always bounded between $1$ and $1/P_d$ and is increasing with $N$. Hence, a sufficient condition for the condition in proposition 3 to be satisfied is to have $\mu_P^{\max}<P_{\mathrm{s}}$ which is clearly satisfied for some $N$, possibly large (by proposition $2$) because $P_{\mathrm{s}}$ can be made arbitrarily close to 1 by increasing the number of secondary nodes. This attracts the attention that the more secondary users the primary shares the channel with, the more benefit for the primary user in terms of his stable throughput. It should be noted that increasing the secondary nodes' transmission power $P_0$ leads to satisfying the condition in proposition 3 for a smaller number of secondary relaying nodes.\\
We also note that one node relaying ($N=1$) always leads to higher primary stable throughput rate if\\
\begin{equation}
\frac{(1-(1-\mu_P^{\max})(1-P_d))P_s}{P_s+(1-\mu_P^{\max})P_d}>\mu_P^{\max}\label{102}
\end{equation}
which can be simplified to:
\begin{equation*}
\mu_P^{\max}=\mathrm{Pr}\left[\frac{P_P|h_{S_PD_P}|^2r_{S_PD_P}^{-\alpha}}{N_0}>\beta_P\right]<P_{\mathrm{s}}=\mathrm{Pr}\left[\frac{P_0|h_i|^2r_0^{-\alpha}}{N_0}>\beta_P\right]
\end{equation*}
We then proceed to calculate the effect of relaying on the secondary nodes. As previously mentioned, secondary nodes might benefit from relaying the primary packets. This can be understood by noting that relaying the primary packets helps the primary to become idle more often, that is, a larger number of idle slots will be available for the secondary nodes. However, portion of that fraction is dedicated for relaying the packets of the primary. If that fraction is smaller than the additional fraction available for the secondary by relaying, a net increase in throughput is realized for the secondary nodes. It should be noted that even if the secondary nodes suffer from some reduction in throughput by relaying, they still achieve some non-zero throughput by accessing the resources of the primary and relaying can then be looked at as the price to pay for accessing the channel. On the other hand, as previously shown, the primary always benefits from relaying when $N$ is sufficiently large and this is the incentive to sharing his resources.\\
The effect of the relaying protocols on the secondary nodes can be found by first rewriting (\ref{15}) as $\lambda_j=\left(1-\frac{\lambda_P}{\mu_P^{\max}}\right)\tilde{\lambda_j}$. A secondary node transmits its own traffic only if the slot is idle, it does not have any primary traffic to relay and no other nodes has any. Note that when a secondary node transmits, it behaves as in the perfect sensing case; thus we obtain the throughput of the $j$th secondary node as:
\begin{equation}
\lambda_j^{\text{relaying}}=\left(1-\frac{\lambda_P}{\mu_P}\right)\left(1-\frac{\lambda_{ext}^P}{\mu_{ext}^P}\right)^N\tilde{\lambda_j}\label{100}\\
\end{equation}
On one hand, the fraction of idle slots available to the secondary nodes is now $\left(1-\frac{\lambda_P}{\mu_P}\right)$ which is larger than the fraction available to the secondary nodes with no relaying given by $\left(1-\frac{\lambda_P}{\mu_P^{\max}}\right)$ because $\mu_P>\mu_P^{\max}$. However, on the other hand, their throughput is reduced by a factor of $\left(1-\frac{\lambda_{ext}^P}{\mu_{ext}^P}\right)^N$ due to the priority given to the relayed primary packets instead of transmitting their own packets. By increasing $N$, the fraction $\left(1-\frac{\lambda_{ext}^P}{\mu_{ext}^P}\right)^N$ decreases due to the priority to transmit primary packets which is the price for the secondary nodes to opportunistically access the channel. We make this precise in the following proposition:\\
\textit{Proposition 4:}\\
The secondary nodes benefit from relaying only if:
\begin{equation}
\left(1-\frac{\lambda_P}{\mu_P^{\max}}\right)<\left(1-\frac{\lambda_P}{\mu_P}\right)\left(1-\frac{\lambda_{ext}^P}{\mu_{ext}^P}\right)^N\label{101}
\end{equation}
In particular, for a system with a single secondary node $(N=1)$, this condition is equivalent to:
\begin{equation*}
\frac{(1-(1-\mu_P^{\max})(1-P_d))P_{\mathrm{s}}}{P_{\mathrm{s}}+(1-\mu_P^{\max})P_d}>\mu_P^{\max}\Leftrightarrow\mu_P^{\max}<P_s
\end{equation*}
\begin{proof}
Follows directly by rewriting (\ref{15}) as $\lambda_j^{\text{no relaying}}=\left(1-\frac{\lambda_P}{\mu_P^{\max}}\right)\tilde{\lambda_j}$ and using (\ref{100}). The case for $N=1$ follows after some algebra by substituting $N=1$ in (\ref{49}) and substituting in (\ref{101}).
\end{proof}
The condition in proposition 4 for the secondary node to benefit from relaying in the case of $N=1$ is identical to the condition in (\ref{102}) for the primary node to benefit from relaying in case of one secondary node. This means that with one secondary node, either both the primary and secondary node benefit from relaying or none of them benefits from relaying.\\
For the imperfect sensing case, although having more secondary users leads to more potential interference with the primary node, it also leads to more benefits of relaying as discussed above. Moreover, it leads to higher opportunities for cooperative channel sensing \cite{Coop_Sensing} leading to more accurate sensing results and thus reducing both false alarm and misdetection probabilities approaching the ideal sensing case discussed above. Quantifying the effect of sensing errors on our cooperative protocol with cooperative sensing is beyond the scope of this paper and is left for future work.\\
Finally, for the special case of Rayleigh fading, the different probabilities and throughputs take the following forms:
\begin{equation}
P_d=\exp\left(\frac{-\beta_PN_0}{\sigma^2P_Pr^{-\alpha}}\right)\label{59}
\end{equation}
\begin{equation}
P_{\mathrm{s}}=\sum_{k=0}^{N-1}\binom{N-1}{k}\frac{P_d^k(1-P_d)^{N-1-k}}{k!}\Gamma\left(k+1,\frac{\beta_PN_0}{P_0r_0^{-\alpha}\sigma_0^2}\right)\label{60}
\end{equation}
and $\Gamma(s,x)$ is the upper incomplete gamma function and can be represented by the integral:
$\Gamma(s,x)=\int_x^\infty t^{s-1}e^{-t}\,dt$.
\begin{equation}
\lambda_j=\left(1-\frac{\lambda_P}{\mu_P}\right)\left(1-\frac{\lambda_{ext}^P}{\mu_{ext}^P}\right)^N\exp\left(\frac{-\beta N_0}{\tilde{\sigma}^2 r_j^{-\alpha}P_0}\right)q\left[1-q\frac{\beta}{1+\beta}\right]^{N-1}\label{61}
\end{equation}
For this special case, one node relaying ($N=1$) always leads to higher primary stable throughput rate if\\
\begin{equation*}
\frac{(1-(1-\mu_P^{\max})(1-P_d))P_s}{P_s+(1-\mu_P^{\max})P_d}>\mu_P^{\max}
\end{equation*}
which can be simplified to:
\begin{align}
&\mu_P^{\max}=\exp\left(\frac{-\beta_PN_0}{P_P\sigma_{S_PD_P}^2r_{S_PD_P}^{-\alpha}}\right)<
P_{\mathrm{s}}=\Gamma\left(1,\frac{\beta_PN_0}{P_0r_0^{-\alpha}\sigma_0^2}\right)=\exp\left(\frac{-\beta_PN_0}{P_0\sigma_0^2r_0^{-\alpha}}\right)\notag\\
&\Leftrightarrow P_0r_0^{-\alpha}\sigma_0^2>P_Pr_{S_PD_P}^{-\alpha}\sigma_{S_PD_P}^2\notag\\
&\Leftrightarrow \textbf{E}[\text{SNR on S-P link}]>\textbf{E}[\text{SNR on P-P link}]\label{62}
\end{align}
In other words, assuming same transmission power for primary and secondary nodes, one node relaying always helps both the primary and the secondary user (by proposition 4) if the channel between secondary source and primary destination is on average better than the channel between the primary source and primary destination.
\section{Numerical Results}
\begin{figure}[h]
\centering
\includegraphics[width=1\textwidth]{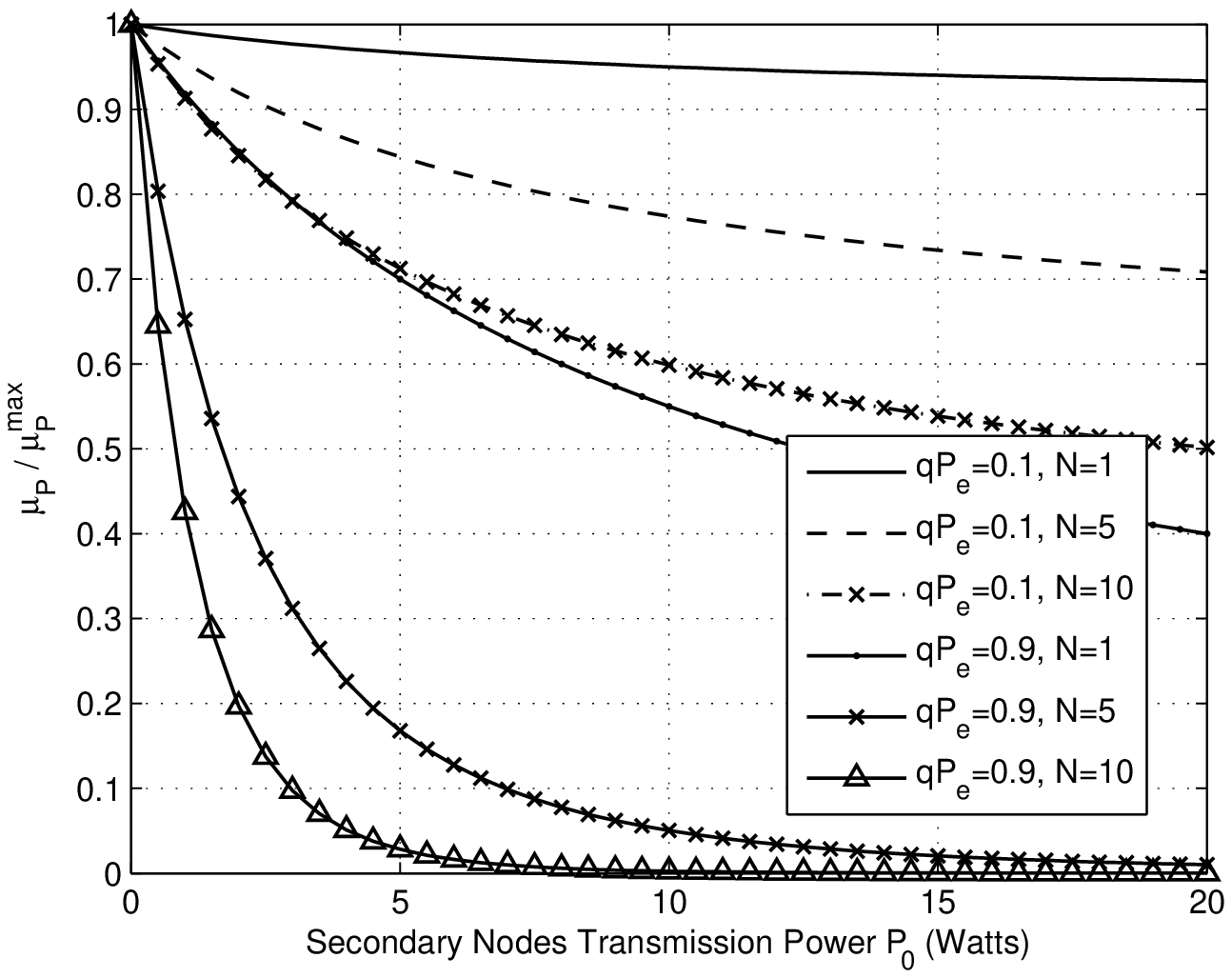}
\caption{Effect of secondary transmission power on primary maximum stable throughput rate}\label{Fig1}
\end{figure}

\begin{figure}[h]
\centering
\includegraphics[width=1\textwidth]{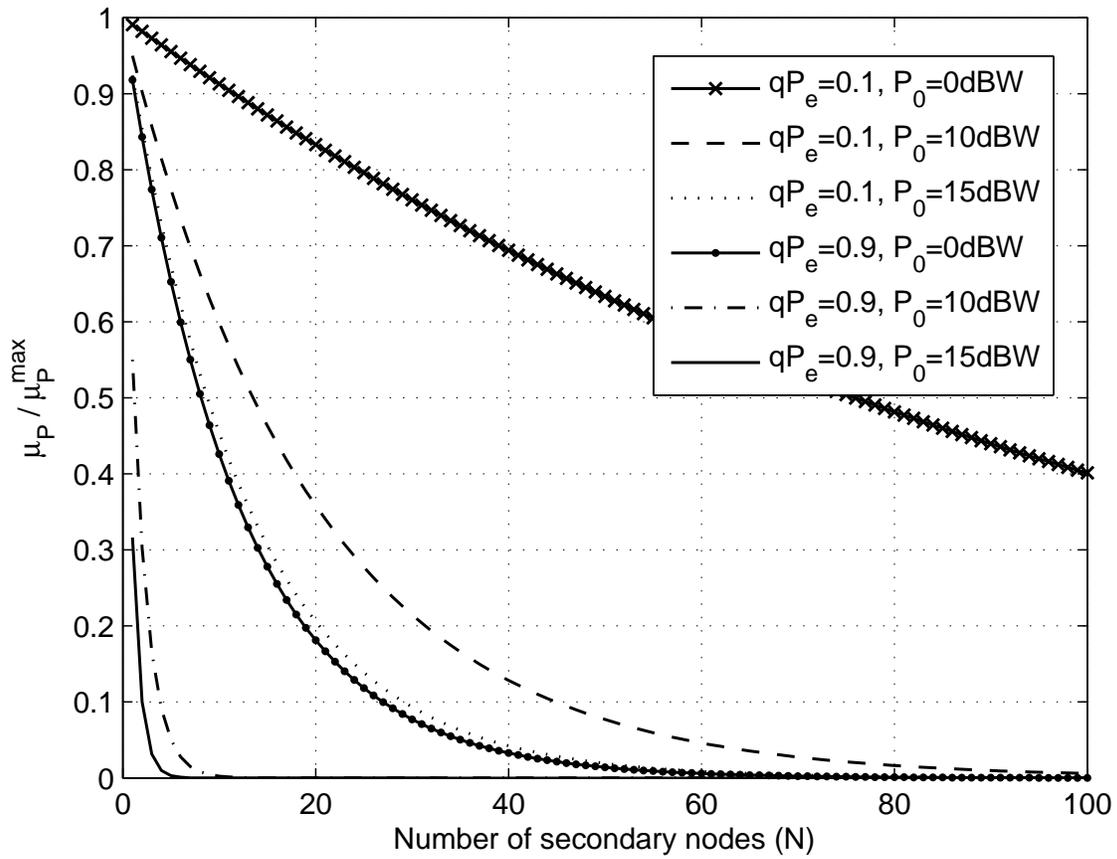}
\caption{Effect of number of secondary nodes on primary maximum stable throughput rate}\label{Fig2}
\end{figure}

\begin{figure}[h]
\centering
\includegraphics[width=1\textwidth]{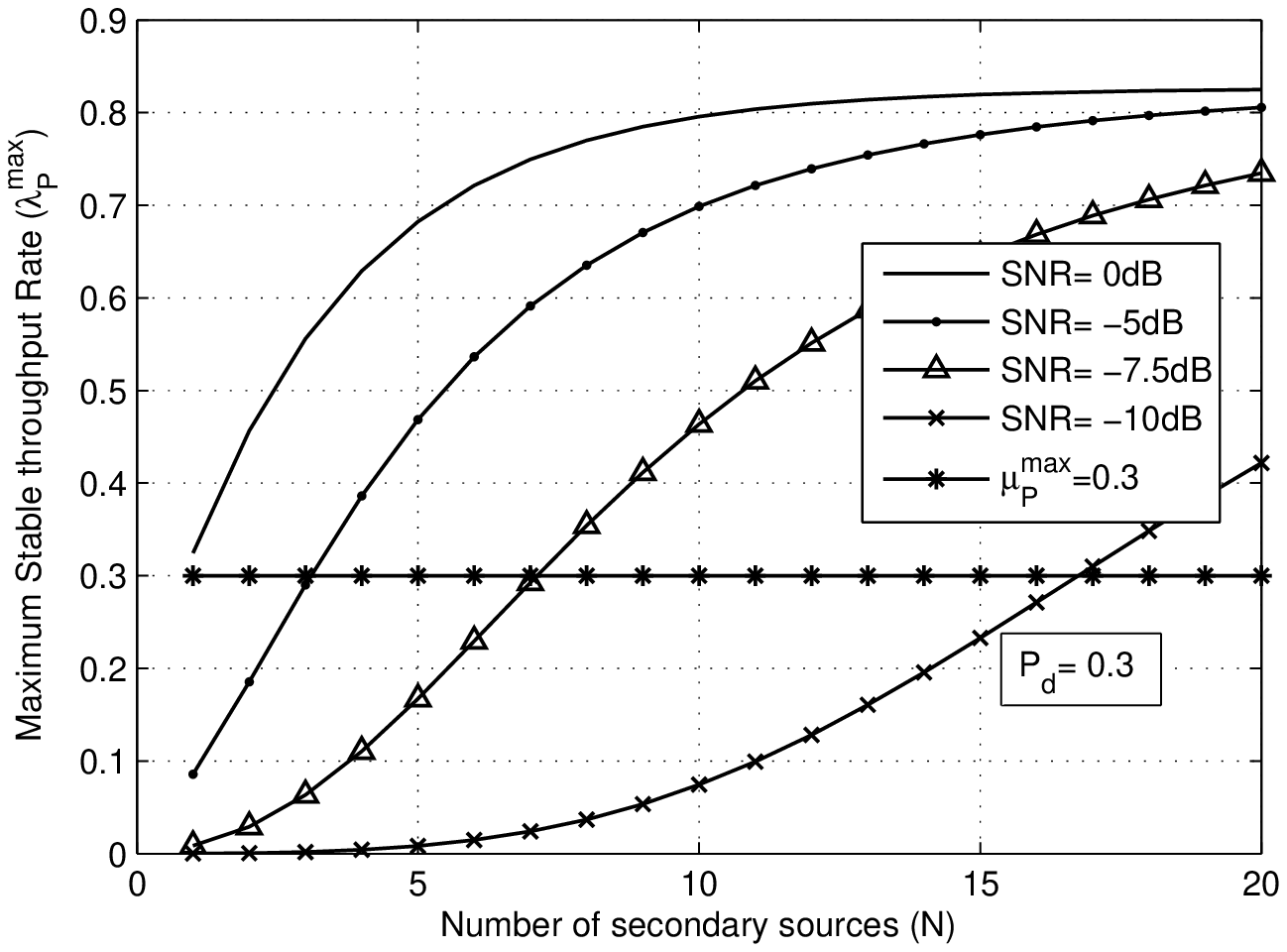}
\caption{Effect of relaying on maximum stable throughput rate $(\lambda_P^{\max})$ for detection probability $P_d=0.3$}\label{Fig5}
\end{figure}

\begin{figure}[h]
\centering
\includegraphics[width=1\textwidth]{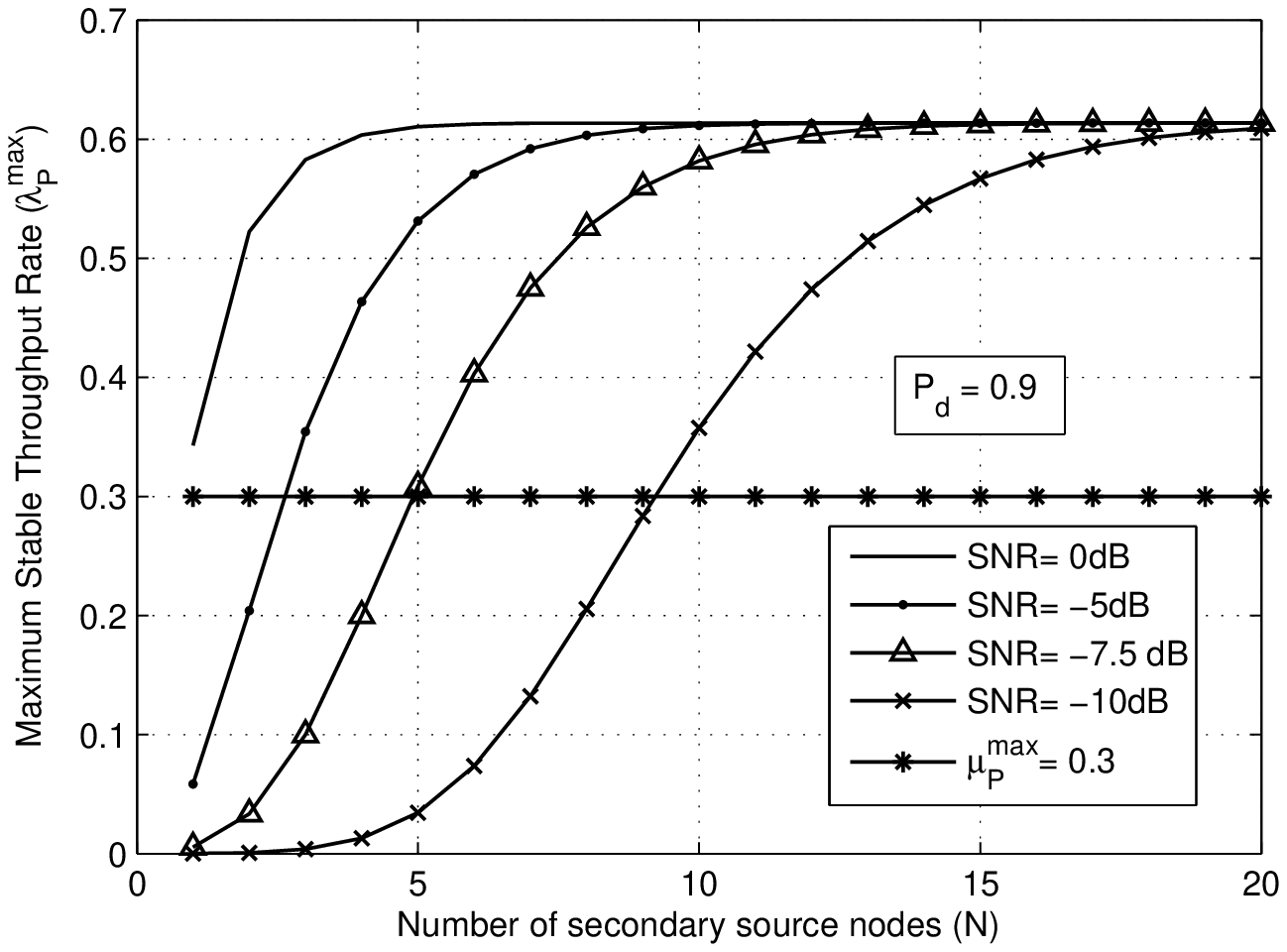}
\caption{Effect of relaying on maximum stable throughput rate $(\lambda_P^{\max})$ for detection probability $P_d=0.9$}\label{Fig6}
\end{figure}

\begin{figure}[h]
\centering
\includegraphics[width=1\textwidth]{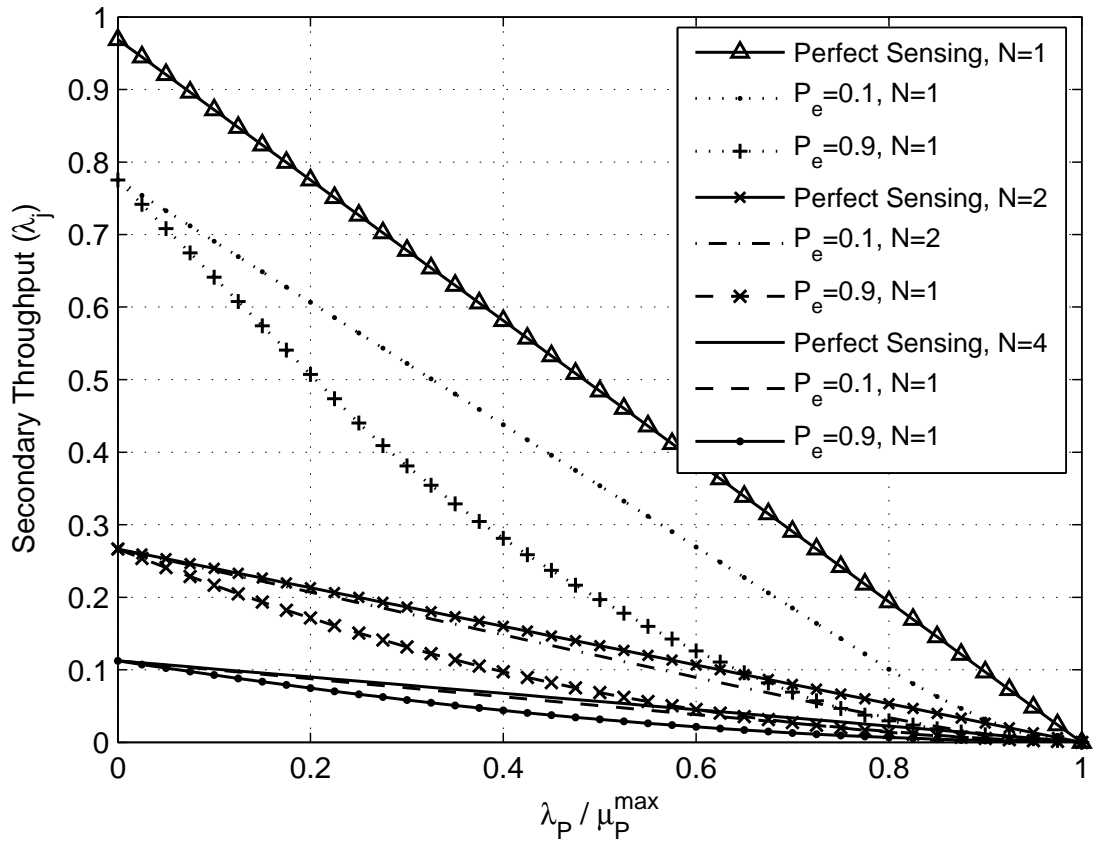}
\caption{Secondary throughput versus primary normalized arrival rate for various values of $P_e$ and $N$. $I=100$ (Case of high interference from the primary).}\label{Fig7}
\end{figure}

\begin{figure}[h]
\centering
\includegraphics[width=1\textwidth]{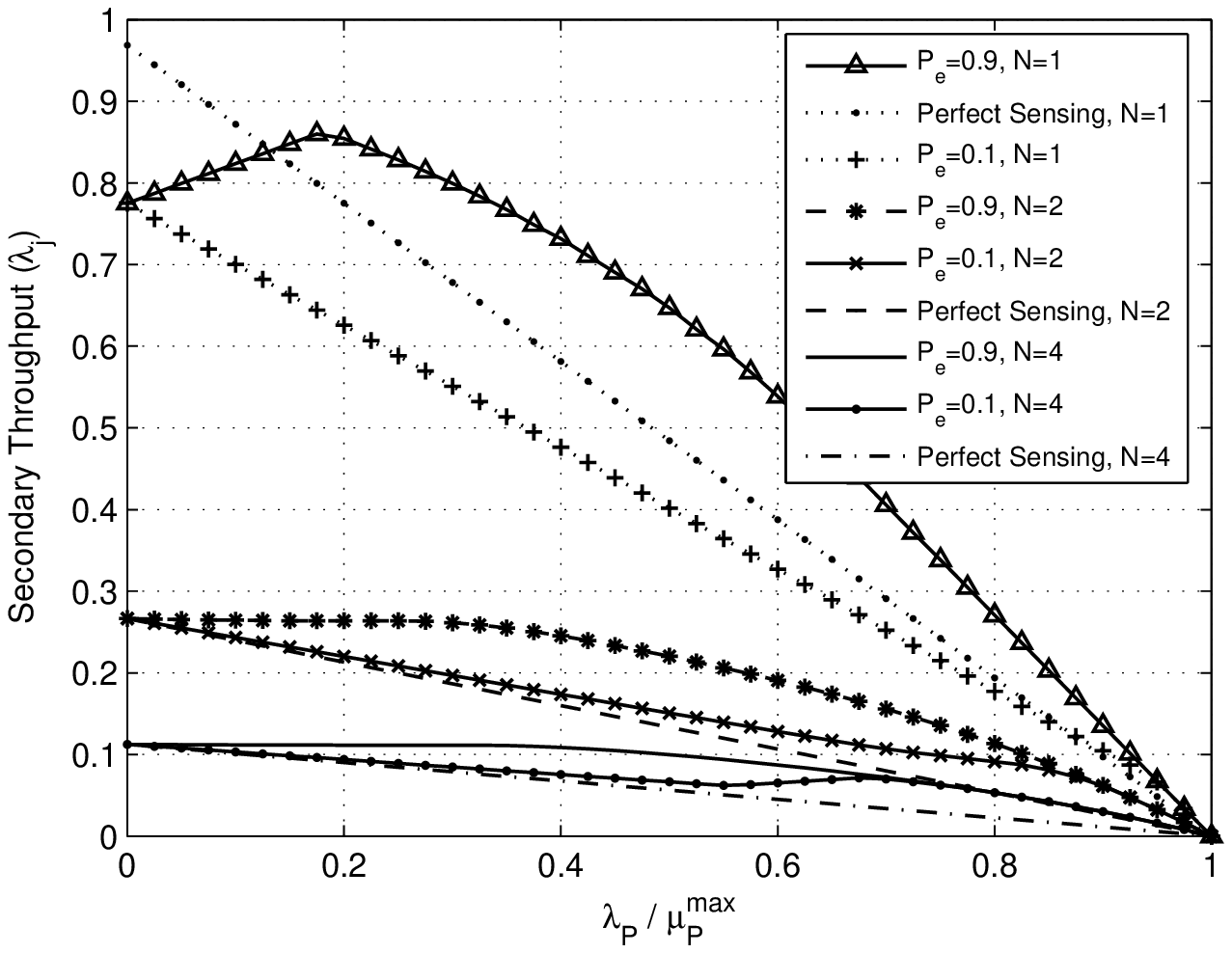}
\caption{Secondary throughput versus primary normalized arrival rate for various values of $P_e$ and $N$. $I=0.1$ (Case of very low interference from the primary).}\label{Fig9}
\end{figure}

In this section, we provide numerical results to illustrate the conclusions drawn analytically. The values of the parameters are chosen based on practical values but also for sake of clarity of presentation. Figures \ref{Fig1} and \ref{Fig2} illustrate the effect of erroneous sensing on the normalized maximum stable throughput of the primary node as given by (\ref{27}). The term $\frac{\sigma_{S_PD_P}^2r_{S_PD_P}^{-\alpha}P_P}{\beta_P r_0^{-\alpha}\sigma_0^2}$ is fixed at value 10dBW throughout figures \ref{Fig1}, \ref{Fig2}. Figure \ref{Fig1} plots the normalized maximum stable throughput of the primary node versus the secondary nodes' transmission power. It shows that $\mu_P$ can severely drop from its perfect sensing value $\mu_P^{\max}$ even for small number of secondary nodes and small values of $qP_e$ and shows that secondary nodes can effectively limit their interference on the primary by controlling their transmission power $P_0$, their channel access probability $q$ or by enhancing the sensing performance to reduce $P_e$ by using better detectors or using cooperative sensing techniques. Figure \ref{Fig2} plots the normalized maximum stable throughput rate at the primary node versus the number of secondary nodes $N$ showing a similar effect. However, it should be noted that as shown in figure \ref{Fig1}, $\lim_{a\to 0}\frac{\mu_P}{\mu_P^{\max}}=\lim_{P_0\to \infty}\frac{\mu_P}{\mu_P^{\max}}=(1-qP_e)^N$, $\lim_{qP_e\to 1}\frac{\mu_P}{\mu_P^{\max}}=[1-\frac{1}{a+1}]^N$ while $\lim_{N\to \infty}\frac{\mu_P}{\mu_P^{\max}}=0$ meaning that for low enough primary arrival rates, controlling the secondary nodes' transmission parameters is not as crucial as controlling the number of secondary transmissions in the system. This motivates the relaying protocol described above which is illustrated in Figures \ref{Fig5} and \ref{Fig6}. Let $SNR=\frac{P_0r_0^{-\alpha}\sigma_0^2}{\beta_PN_0}$. Figures \ref{Fig5} and \ref{Fig6} plot the maximum stable throughput rate at the primary node ($\lambda_P^{\max}$) as given by (\ref{57}) versus the number of secondary nodes for different SNR values for the case of Rayleigh fading (\ref{59}), (\ref{60}). With no relaying, $\lambda_P^{\max}=\mu_P^{\max}$ and is shown by the horizontal line at $0.3$. It is clear that regardless of the parameters' values, sufficiently large $N$ always outperforms the non-relaying case and with higher SNR, a smaller number of secondary nodes is needed to outperform non-relaying. We also note that at SNR = 0 dB, even one node relaying leads to better performance than non-relaying case. Figures \ref{Fig7} and \ref{Fig9} show the secondary nodes maximum throughput (optimized over $q$ and $P_0$) versus the normalized average arrival rate at the primary node $\lambda_P/\mu_P^{\max}$ for different values of $I=\frac{P_Pr_{S_PD_j}^{-\alpha}\beta\sigma_{S_PD_j}^2}{r_j^{-\alpha}\tilde{\sigma}^2}$ in both perfect and erroneous sensing cases as given by equations (\ref{15}) and (\ref{45}) respectively. Note that for each value of $\lambda_P/\mu_P^{\max}$, the feasible set of $(q,P_0)$ may be different to ensure the protection of the primary node as in (\ref{30}) and (\ref{31}). The value of the secondary threshold $\beta$ is fixed at 10, $\frac{\beta N_0}{\tilde{\sigma}^2r_j^{-\alpha}}=-5$ dBW,$\frac{\sigma_{S_PD_P}^2P_Pr_{S_PD_P}^{-\alpha}}{\sigma_0^2\beta_Pr_0^{-\alpha}}=0$ dBW and $P_f=0.2$. We also impose a maximum possible value on $P_0$ equal to $10$dBW which is typically a constraint imposed by the hardware. Figure \ref{Fig7} shows the secondary throughput for $I=100$ which is the case where the primary node exerts high interference on the secondary nodes. In this case, as shown, perfect sensing leads to a higher throughput compared with imperfect sensing. Furthermore, the throughput $\lambda_j$ decreases with increasing the error probability $P_e$ because of the decrease of idle slots (primary interference free) in spite of the increase of the busy slots suffering from high primary interference and thus cannot balance the reduction of the relatively high throughput acquired in idle slots. Figure \ref{Fig9} shows the case of $I=0.1$ which is the case of very low interference from the primary. In contrast with figure \ref{Fig7} in which the secondary throughput decreases with increasing $\lambda_P/\mu_P^{\max}$, for some parameters values (for instance, $N=1, P_e=0.9$) secondary throughput increases with $\lambda_P/\mu_P^{\max}$. Moreover, except for $N=1$, incorrect sensing leads to a higher throughput than perfect sensing, and increasing $P_e$ leads to an increase in throughput. Hence, in this case, although increasing $P_e$ might harm the primary node, secondary nodes benefit in terms of their throughputs. This is due to the increase of the  opportunities in which secondary nodes access the channel as the fraction of busy slots suffering from low primary interference increases. This case is appealing to the secondary nodes if the primary arrival rate is low enough allowing them to increase $P_e$ to the level which does not affect the primary node stability as discussed previously.
\section{Conclusion and Future Work}
In this paper, we studied the effect of the number of secondary nodes and their transmission parameters on the stable throughput of the primary user as well as on the secondary's throughputs in both perfect and imperfect sensing cases. It was shown that secondary transmission parameters (power and channel access probabilities) must depend on the arrival rate of the primary to ensure the primary's protection. If the primary user's arrival rate is less than some calculated finite value, there is no need for controlling their parameters, otherwise, secondary nodes have to control their transmission parameters to limit their interference on the primary and avoid affecting its stability. The number of secondary users can be a benefit or a hindrance. If the secondary nodes do not relay the primary unsuccessful packets, their presence is a harm for the primary as it reduces its maximum stable throughput. However, if the secondary nodes are forced to relay the primary, then the primary always benefits from having many nodes relaying its packets and secondary nodes might benefit from having access to a larger fraction of idle slots. This observation reveals that with appropriate relaying protocols, cognitive radio technology is appealing for licensed users to share their resources with other unlicensed users. Future work includes the extension of the analysis to more complicated topologies such as line and grid networks as well as the analysis of our relaying protocol in the imperfect sensing case. Also, the extension to the case of cooperative sensing instead of single node sensing will lead to a more practical model, and in this case, the study of the effect of sensing overhead on the throughput will be critical.
\appendices
\section{Proof of (\ref{7}) and (\ref{14})}
We use the following lemma in the proof:\\
\textit{Lemma A.1:}\\
Let $X_i\sim\exp\left(\theta_i\right)$ be independent random variables, then the probability density function of the sum $Z=\sum_{i=1}^NX_i$ is given by:
\begin{equation}
f_Z(z)=\sum_{i=1}^N\left(\prod_{\substack{j=1\\j\ne i}}^N\frac{\theta_j\theta_i}{\theta_j-\theta_i}\right)\exp(-\theta_iz)\label{63}
\end{equation}
If the random variables $X_i$ are also identically distributed, i.e., $X_i\sim\exp(-\theta)$ for all $i$, then the sum $Z=\sum_{i=1}^NX_i$ has pdf given by the Erlang distribution:
\begin{equation}
f_Z(z)=\theta\exp(-\theta z)\frac{\left(\theta z\right)^{N-1}}{(N-1)!}\label{64}
\end{equation}
The proof is by induction. Refer to \cite{Ross} for details.\\
Hence, for the case of asymmetric configuration:\\
\begin{align}
&\mathrm{Pr}\left[\frac{P_{S_j}|h_{S_jD_j}|^2r_{S_jD_j}^{-\alpha}}{N_0+\sum_{k\in \mathcal{T},k\ne j}P_k|h_{S_kD_j}|^2r_{S_kD_j}^{-\alpha}}>\beta_{j}\right]=\mathrm{Pr}\left[|h_{S_jD_j}|^2>\frac{\beta_jN_0}{P_jr_{S_jD_j}^{-\alpha}}+Z\right]
\notag\\&=\int_0^\infty\mathrm{Pr}\left[|h_{S_jD_j}|^2>\frac{\beta_jN_0}{P_jr_{S_jD_j}^{-\alpha}}+z\right]\sum_{\substack{k\in\mathcal{T}\\k\ne j}}\prod_{l\ne k}\frac{\theta_l\theta_{k}}{\theta_l-\theta_k}\exp(-\theta_{k}z)\,dz\notag\\ &=\exp\left(\frac{-N_0\beta_j}{\sigma_{S_jD_j}^2P_jr_{S_jD_j}^{-\alpha}}\right)\sum_{\substack{k\in\mathcal{T}\\k\ne j}}\left(\prod_{l\ne k}\frac{\theta_l\theta_k}{\theta_l-\theta_k}\right)\frac{1}{\left(\theta_k+1/\sigma_{S_jD_j}^2\right)}\label{65}
\end{align}
where: $\theta_k=\frac{P_jr_{S_jD_j}^{-\alpha}}{P_kr_{S_kD_j}^{-\alpha}\sigma_{S_kD_j}^2\beta_j}$.\\
For the case of symmetric network:
\begin{align}
&\mathrm{Pr}\left[\frac{P_0|h_{S_jD_j}|^2r_j^{-\alpha}}{N_0+\sum_{m=1}^k{P_0|h_m|^2r_j^{-\alpha}}}>\beta\right]=
\mathrm{Pr}\left[|h_{S_jD_j}|^2>\frac{\beta N_0}{P_0r_j^{-\alpha}}+\beta\sum_{l=1}^k|h_{S_lD_j}|^2\right]\notag\\
&=\int_0^\infty\mathrm{Pr}\left[|h_{S_jD_j}|^2>\frac{\beta N_0}{P_0r_j^{-\alpha}}+\beta z\right]\frac{\exp(-z/{\tilde{\sigma}}^2)(z/{\tilde{\sigma}}^2)^{k-1}}{{\tilde{\sigma}}^2(k-1)!}\,dz=\exp\left(\frac{-\beta N_0}{{\tilde{\sigma}}^2P_0r_j^{-\alpha}}\right)\frac{1}{(1+\beta)^k}\label{66}
\end{align}
where we have used that $\int_0^\infty x^{k-1}\exp(-x)\,dx=(k-1)!$, for $k$ integer.
\section{Proof of Relaying Protocol in the General Asymmetric Case}
We use the following Lemma in the proof:\\
\textit{Lemma B.1:}\\
If $X_1,X_2,...,X_N$ are positive random variables, then for every $c>0$:
\begin{equation}
\mathrm{Pr}\left[\sum_{i=1}^{k+1}X_i>c\right]>\mathrm{Pr}\left[\sum_{i=1}^{k}X_i>c\right],\text{ }k\in\{1,2,...,N-1\}\label{67}
\end{equation}
\textit{Proof:} Follows by induction and is omitted for space limitations.\\
Let the flat fading coefficients between the primary source and the $j$th secondary source nodes $h_{S_PS_j}$. The probability that the $j$th secondary node is able to successfully decode the primary packet is then given by:
\begin{equation}
P_d^{(j)}=\mathrm{Pr}\left[\frac{P_P|h_{S_PS_j}|^2r_{S_PS_j}^{-\alpha}}{N_0}>\beta_P\right],\mbox{  }\{1,2,...,N\}\label{70}
\end{equation}
Let $\mathcal{S}=\{1,2,...,N\}$ be the set of secondary nodes.\\
The probability that some set of the $N$ secondary nodes is able to successfully decode the primary packet is given by:
\begin{equation}
P_d^{(N)}=1-\prod_{k\in\mathcal{S}}\left(1-P_d^{(k)}\right)\label{71}
\end{equation}
where the superscript $N$ denotes the number of secondary nodes in the system.\\
According to the relaying protocol, the primary node is served either when the primary destination can successfully decode the packet or when the primary destination cannot and one or more secondary source nodes can. Hence, the average service rate of the primary node is given by:
\begin{equation}
\mu_P=\mu_P^{\max}+\left(1-\mu_P^{\max}\right)P_d^{(N)}=P_d^{(N)}+\left(1-P_d^{(N)}\right)\mu_P^{\max}\label{72}
\end{equation}
which is clearly greater than $\mu_P^{\max}$.\\
If the relaying queues of the secondary nodes are stable (i.e. they empty infinitely often), then this ensures that the primary's packets which are successfully decoded by the secondary nodes will eventually reach the primary destination.\\
The average arrival rate to the $j$th secondary node relaying queue is given by:
\begin{equation}
\lambda_j^{ext}=\left(\frac{\lambda_P}{\mu_P}\right)\left(1-\mu_P^{\max}\right)P_d^{(j)}\label{73}
\end{equation}
The average service rate of the $j$th secondary relaying queue is given by: $\left(1-\frac{\lambda_P}{\mu_P}\right)P_s^{(j)}$.\\
where:
\begin{equation}
P_s^{(j)}=\sum_{T\subseteq\{1,2,..,N\}\setminus\{j\}}\left[\prod_{k\in T}P_d^{(k)}\prod_{k\in\{1,2,...,N\}\setminus\{T\bigcup j\}}\left(1-P_d^{(k)}\right)\right]\mathrm{Pr}\left[\frac{\sum_{i\in \{T\bigcup j\}}P_i|h_{S_iD_P}|^2r_{S_iD_P}^{-\alpha}}{N_0}>\beta_P\right]\label{74}
\end{equation}
For the stability of all the secondary nodes' relaying queues, we must have for all $j=1,2,...,N$:
\begin{equation}
\left(\frac{\lambda_P}{\mu_P}\right)\left(1-\mu_P^{\max}\right)P_d^{(j)}<\left(1-\frac{\lambda_P}{\mu_P}\right)P_s^{(j)}\label{75}
\end{equation}
which is equivalent to:
\begin{equation}
\lambda_P<\min_{1\leq j\leq N}\left\{\frac{\mu_PP_s^{(j)}}{P_s^{(j)}+\left(1-\mu_P^{\max}\right)P_d^{(j)}}\right\}\label{76}
\end{equation}
\textit{Lemma B.2:}\\
$P_s^{(j)}$ as given in (\ref{74}) is monotone increasing with $N$ and converges to $1$ as $N\to\infty$.
\begin{proof}\\
The event $\left\{\frac{\sum_{i\in \{T\bigcup j\}}P_i|h_{S_iD_P}|^2r_{S_iD_P}^{-\alpha}}{N_0}>\beta_P\right\}$ implies the event $\left\{\frac{\sum_{i\in \{1,2,...,N\}}P_i|h_{S_iD_P}|^2r_{S_iD_P}^{-\alpha}}{N_0}>\beta_P\right\}$, hence:
\begin{equation}
\mathrm{Pr}\left[\frac{\sum_{i\in \{T\bigcup j\}}P_i|h_{S_iD_P}|^2r_{S_iD_P}^{-\alpha}}{N_0}>\beta_P\right]\leq\mathrm{Pr}\left[\frac{\sum_{i\in \{1,2,...,N\}}P_i|h_{S_iD_P}|^2r_{S_iD_P}^{-\alpha}}{N_0}>\beta_P\right]\label{77}
\end{equation}
Using that $\sum_{T\subseteq\{1,2,..,N\}\setminus\{j\}}\left[\prod_{k\in T}P_d^{(k)}\prod_{k\in\{1,2,...,N\}\setminus\{T\bigcup j\}}\left(1-P_d^{(k)}\right)\right]=1$, we get that $P_s^{(j)}$ is upper bounded by $\mathrm{Pr}\left[\frac{\sum_{i\in \{1,2,...,N\}}P_i|h_{S_iD_P}|^2r_{S_iD_P}^{-\alpha}}{N_0}>\beta_P\right]$.\\
By Lemma B.1, this bound converges to $1$ as $N\to\infty$. We need to show that $P_s^{(j)}$ is monotone increasing with $N$ and hence must converge to $1$ as $N\to\infty$ by monotone convergence theorem.\\
To show monotonicity, consider:
\begin{align}
&P_s^{(j)}(N)=\sum_{T\subseteq\mathcal{S}\setminus\{j\}}\left[\prod_{k\in T}P_d^{(k)}\prod_{k\in\mathcal{S}\setminus\{T\bigcup j\}}\left(1-P_d^{(k)}\right)\right]\mathrm{Pr}\left[\frac{\sum_{i\in \{T\bigcup j\}}P_i|h_{S_iD_P}|^2r_{S_iD_P}^{-\alpha}}{N_0}>\beta_P\right]\notag\\
&P_s^{(j)}(N+1)=\sum_{T\subseteq\mathcal{S}\bigcup\{N+1\}\setminus\{j\}}\left[\prod_{k\in T}P_d^{(k)}\prod_{k\in\mathcal{S}\bigcup\{N+1\}\setminus\{T\bigcup j\}}\left(1-P_d^{(k)}\right)\right]\mathrm{Pr}\left[\frac{\sum_{i\in\{T\bigcup j\}}P_i|h_{S_iD_P}|^2r_{S_iD_P}^{-\alpha}}{N_0}>\beta_P\right]\label{78}
\end{align}
The summation in $P_s^{(j)}(N+1)$ has twice as many terms as the summation in $P_s^{(j)}(N)$. Specifically, each set in $P_s^{(j)}(N)$ exists in $P_s^{(j)}(N+1)$ as well as the same set union the set $\{N+1\}$.\\
Let $\mathcal{M}$ be the set of all sets in $P_s^{(j)}(N)$, then:
\begin{equation}
P_s^{(j)}(N)=\sum_{m\in\mathcal{M}}\left[\prod_{k\in m}P_d^{(k)}\prod_{k\in\{1,2,...,N\}\setminus\{m\bigcup j\}}\left(1-P_d^{(k)}\right)\right]\mathrm{Pr}\left[\frac{\sum_{i\in \{m\bigcup j\}}P_i|h_{S_iD_P}|^2r_{S_iD_P}^{-\alpha}}{N_0}>\beta_P\right]
\end{equation}
\begin{align}
&P_s^{(j)}(N+1)=\sum_{m\subseteq\mathcal{M}}\left[\prod_{k\in m}P_d^{(k)}\prod_{k\in\{1,2,...,N+1\}\setminus\{m\bigcup j\}}\left(1-P_d^{(k)}\right)\right]\mathrm{Pr}\left[\frac{\sum_{i\in\{m\bigcup j\}}P_i|h_{S_iD_P}|^2r_{S_iD_P}^{-\alpha}}{N_0}>\beta_P\right]\notag\\
&+\sum_{m\subseteq\mathcal{M}}\left[\prod_{k\in \{m\bigcup\{N+1\}\}}P_d^{(k)}\prod_{k\in\{1,2,...,N\}\setminus\{m\bigcup j\}}\left(1-P_d^{(k)}\right)\right]\mathrm{Pr}\left[\frac{\sum_{i\in\{m\bigcup j\bigcup\{N+1\}\}}P_i|h_{S_iD_P}|^2r_{S_iD_P}^{-\alpha}}{N_0}>\beta_P\right]\label{79}
\end{align}
By using that
\begin{equation}
\mathrm{Pr}\left[\frac{\sum_{i\in\{m\bigcup j\bigcup\{N+1\}\}}P_i|h_{S_iD_P}|^2r_{S_iD_P}^{-\alpha}}{N_0}>\beta_P\right]>\mathrm{Pr}\left[\frac{\sum_{i\in\{m\bigcup j\}}P_i|h_{S_iD_P}|^2r_{S_iD_P}^{-\alpha}}{N_0}>\beta_P\right]
\end{equation}
we get:
\begin{align}
&P_s^{(j)}(N+1)\geq\sum_{m\subseteq\mathcal{M}}\left[P_d^{(N+1)}\prod_{k\in m}P_d^{(k)}\prod_{k\in\{1,2,...,N\}\setminus\{m\bigcup j\}}\left(1-P_d^{(k)}\right)\right.+\notag\\
&\left.\prod_{k\in m}P_d^{(k)}\prod_{k\in\{1,2,...,N\}\setminus\{m\bigcup j\}}\left(1-P_d^{(k)}\right)\left(1-P_d^{(N+1)}\right)\right]\mathrm{Pr}\left[\frac{\sum_{i\in\{m\bigcup j\}}P_i|h_{S_iD_P}|^2r_{S_iD_P}^{-\alpha}}{N_0}>\beta_P\right]\notag\\
&=\sum_{m\subseteq\mathcal{M}}\left[\prod_{k\in m}P_d^{(k)}\prod_{k\in\{1,2,...,N\}\setminus\{m\bigcup j\}}\left(1-P_d^{(k)}\right)\right]\mathrm{Pr}\left[\frac{\sum_{i\in\{m\bigcup j\}}P_i|h_{S_iD_P}|^2r_{S_iD_P}^{-\alpha}}{N_0}>\beta_P\right]=P_s^{(j)}(N)\label{80}
\end{align}
\end{proof}
Hence the sequence $P_s^{(j)}(N)$ is monotone increasing in $N$ and upper bounded by $1$, hence, converges to $1$ as $N\to\infty$ for all $j\in\{1,2,...,N\}$.\\
The stability condition of the primary node in case of relaying given by (\ref{76}) can be approximated
if $N$ is sufficiently large ($P_s^{(j)}\to 1$) by:
\begin{equation}
\lambda_P<\frac{\mu_P}{1+\left(1-\mu_P^{\max}\right)\max_{1\leq j\leq N}{\{P_d^{(j)}\}}}\label{81}
\end{equation}
The maximum stable throughput rate in equation (\ref{81}) is larger than $\mu_P^{\max}$ only if $\mu_P^{\max}<\frac{P_d^{(N)}}{\max_{1\leq j\leq N}\{P_d^{(j)}\}}$.\\
Finally, by noting that $P_d^{(N)}=1-\prod_{k\in\{1,2,...,N\}}\left(1-P_d^{(k)}\right)\geq\max_{1\leq j\leq N}\{P_d^{(j)}\}$, this condition is always satisfied for sufficiently large $N$ and hence, relaying always increases the stable throughput of the primary for sufficiently large $N$.
\section{Proof of Proposition 2}
We proceed by showing that $P_s$ is strictly increasing with $N$.
\begin{align}
\psi(N)=P_s&=\sum_{k=0}^{N-1}\binom{N-1}{k}P_d^k(1-P_d)^{N-1-k}\mathrm{Pr}\left[\sum_{i=1}^{k+1}|h_i|^2>c\right]\notag\\
&=\sum_{k=0}^{N-1}\binom{N-1}{k}P_d^k(1-P_d)^{N-1-k}\phi(k)\label{68}
\end{align}
\begin{align}
\psi(N+1)-\psi(N)&=\sum_{k=0}^{N}\binom{N}{k}P_d^k(1-P_d)^{N-k}\phi(k)-\sum_{k=0}^{N-1}\binom{N-1}{k}P_d^k(1-P_d)^{N-1-k}\phi(k)\notag\\
&=P_d^N\phi(N)+\sum_{k=0}^{N-1}\binom{N-1}{k}P_d^k(1-P_d)^{N-1-k}\left[\frac{k-NP_d}{N-k}\right]\phi(k)\notag\\
&=P_d^N\phi(N)+\sum_{k=0}^{\lfloor NP_d\rfloor}\binom{N-1}{k}P_d^k(1-P_d)^{N-1-k}\left[\frac{k-NP_d}{N-k}\right]\phi(k)\notag\\&+\sum_{k=\lceil NP_d\rceil}^{N-1}\binom{N-1}{k}P_d^k(1-P_d)^{N-1-k}\left[\frac{k-NP_d}{N-k}\right]\phi(k)\notag\\
&>P_d^N\phi(N)+\sum_{k=0}^{N-1}\binom{N-1}{k}P_d^k(1-P_d)^{N-1-k}\left[\frac{k-NP_d}{N-k}\right]\phi(\lceil NP_d\rceil)\notag\\
&=P_d^N(\phi(N)-\phi(\lceil NP_d\rceil))>0\label{69}
\end{align}
where we have used the fact that $\phi(N)>\phi(N-1)>...>\phi(1)$ which follows by lemma B.1\\
Hence, $P_s$ is a monotone increasing sequence with supremum equal to one and hence converges to one by the monotone convergence theorem.
\bibliographystyle{unsrt}
\bibliography{Cognitive_MA}

\end{document}